\documentclass[journal=jctcce,manuscript=article,layout=traditional]{achemso}

\usepackage{graphicx}
\usepackage{caption}
\usepackage{subcaption}
\usepackage{bm}
\usepackage{braket}
\usepackage[colorlinks=true,breaklinks=true,linkcolor=red,citecolor=magenta,urlcolor=blue]{hyperref}
\usepackage{comment}
\usepackage{amsmath}
\usepackage{amsfonts,amssymb}
\usepackage{multirow}
\usepackage{multicol}
\usepackage{xcolor}

\usepackage[utf8]{inputenc}
\usepackage[T1]{fontenc}
\usepackage{qcircuit}
\usepackage{float}
\usepackage{verbatim}

\title{\Large{Selecting optimal unrestricted Hartree-Fock trial wavefunctions for phaseless auxiliary-field quantum Monte Carlo:  Accuracy and limitations in modeling three iron-sulfur clusters}}

\author{Don Danilov}
\affiliation{Department of Chemistry, Rice University, Houston, TX 77005-1892, USA}
\altaffiliation{\small These authors contributed equally to this work}
\author{Brad Ganoe}
\affiliation{Department of Chemistry, Rice University, Houston, TX 77005-1892, USA}
\altaffiliation{\small These authors contributed equally to this work}
\author{Leon Otis}
\affiliation{Department of Chemistry, Rice University, Houston, TX 77005-1892, USA}
\author{Zhi Gong}
\affiliation{Department of Chemistry, Rice University, Houston, TX 77005-1892, USA}
\author{Zixiang Lu}
\affiliation{University of Illinois Urbana-Champaign, Urbana, IL 61801, USA}
\author{James Shee}
\affiliation{Department of Chemistry, Rice University, Houston, TX 77005-1892, USA}
\alsoaffiliation{Department of Physics and Astronomy, Rice University, Houston, TX 77005-1892, USA}
\email{james.shee@rice.edu}

\begin{document}

\begin{abstract}
Phaseless auxiliary-field quantum Monte Carlo (ph-AFQMC) has emerged as a promising electronic structure method for correlated electronic systems.  However, the quality of its predictions depends critically on the choice of trial wavefunction, and it is not obvious how to make an optimal choice especially for strongly correlated states of large systems.  
Mean-field wavefunctions are compelling trial wavefunction candidates as they map directly to chemical concepts and can be obtained with $\mathcal{O}(N^4)$ cost. Yet in the strongly correlated regime one faces a symmetry dilemma and the existence of multiple nearly-degenerate solutions. 
In this work we investigate active space models of [2Fe--2S]$^{2+}$, mixed-valent [4Fe--4S]$^{2+}$, and [4Fe--4S]$^{4+}$ and explore the sensitivity of ph-AFQMC to the choice of unrestricted Hartree-Fock trial wavefunction.  We find that chemical properties and physical symmetries, rather than the variational energy, ought to guide the choice of mean-field trial for ph-AFQMC (or reference state for coupled cluster models), and show that surprisingly accurate ground-state energies for these systems can be obtained.  However, in all cases we find a rapidly vanishing overlap between the stochastic wavefunction and the UHF trial, indicating that the trials are suboptimal importance functions.  By analogy to a similar situation in the stretched helium dimer cation, we show how this sampling bias pushes ph-AFQMC towards artificially negative energies, which evidently can be compensated for by the phaseless bias in certain cases.

\end{abstract}

\maketitle

\section{Introduction}

It has been known since the 1950s\cite{bezkorovainy_iron-sulfur_1980} that polynuclear transition metal clusters have significant biological relevance across a wide range of redox reactions.\cite{johnson_structure_2005, gupta_ironsulfur_2020, tanifuji_metalsulfur_2020, read_mitochondrial_2021} 
Prominent examples of their electrocatalytic roles include oxygen evolution at the tetra-manganese oxygen evolving complex (OEC) in Photosystem II,\cite{askerka2017o2,yano2014mn4ca} electron transfer at di- and tetra-nuclear iron-sulfur clusters,\cite{beinert_iron-sulfur_1997} and nitrogen reduction via the FeMo cofactor in nitrogenase enzymes.\cite{hoffman2014mechanism}
Alongside advances in experimental techniques, tremendous efforts have been made using computational quantum chemistry methods such as density functional theory (DFT) and related broken-symmetry approaches to elucidate magnetic and structural properties along with reaction mechanisms.\cite{noodleman1992density,krewald2013magnetic,krewald2015metal,siegbahn2013water,blomberg2014quantum,siegbahn2019mechanism,sproviero2008computational,cox2013biological,benediktsson2022analysis}

However, it is also known that polynuclear transition metal clusters pose a challenge for approximate electronic structure models based on the mean-field formalism, including DFT.  The presence of multiple entangled spins can  give rise to strong electron correlation\cite{ganoe_notion_2024} in catalytically-relevant low-energy states, which requires multi-configurational wavefunctions\cite{shee2021revealing} (though many can be qualitatively captured by one or a relatively small number of Configuration State Functions\cite{izsak_measuring_2023,li2020compression,leyser2024restricted, marti-dafcik_spin_2025,song_spin-adapted_2026}).
In this regime, a growing number of studies have found that DFT is very sensitive to the choice of exchange-correlation functional\cite{cao2019extremely,shee_achieving_2019,zhai2023multireference}, which motivates the need to develop and employ more reliable \emph{ab initio}  many-body techniques. 

Chan and coworkers performed the first many-body wavefunction studies of the OEC\cite{kurashige2013entangled} and di- and tetra-nuclear Fe-S clusters
\cite{sharma_low-energy_2014} using the Density Matrix Renormalization Group (DMRG).  More recently, DMRG studies of other iron-sulfur clusters in nitrogenase, including the P-cluster\cite{li_electronic_2019} and FeMoco,\cite{zhai2026classical} have been reported.  These studies provide \emph{ab initio} descriptions of the many-body wavefunctions and properties such as spin-spin correlations and natural orbital occupation numbers, albeit in reduced active space models containing only the Fe 3$d$ orbitals and S 3$p$ orbitals.  Systematic extrapolation of the bond-dimension parameter enabled near-exact ground-state energy estimates.  Among their major conclusions is that an effective interacting spin Hamiltonian does not always capture the low-energy physics, which has important implications regarding the validity of widely-used strategies such as broken-symmetry DFT that are based on Ising/Heisenberg spin physics.  These Fe-S active space models have become popular targets for quantum computation/simulation due to their complexity and relatively small system sizes, especially in the 25-100 logical qubit regime.\cite{alexeev_perspective_2025, li_electronic_2019, reiher_elucidating_2017,robledo-moreno_chemistry_2025,lee_evaluating_2023}.
There is also an impressive body of work by Li Manni and coworkers, which pioneers the use of spin-adapted multi-configurational wavefunctions, realized via compression techniques\cite{li2020compression,song2025genetic} or a stochastic FCIQMC approach\cite{dobrautz2021spin}, to polynuclear transition metal clusters such as the oxygen-evolving complex\cite{manni2021modeling} and iron-sulfur clusters.\cite{li2021resolution,dobrautz2021spin} 
Multi-reference wavefunction methods have also been combined with more approximate treatment of the protein environment in, e.g., a recent study by Pantazis and coworkers.\cite{drosou2024combined}

Another emerging electronic structure method, which has the ability to scale up to hundreds and even thousands of orbitals with typical trial wavefunctions is phaseless auxiliary-field quantum Monte Carlo (ph-AFQMC).\cite{zhang2003quantum,motta_ab_2018,shi2021some,lee_twenty_2022} Initially developed in the condensed matter physics community in the context of lattice models, ph-AFQMC has undergone intense development in the quantum chemistry community in recent years.  While some consensus is emerging regarding the requisite polynomial-cost trial wavefunctions for the accurate description of weakly correlated molecular systems -- including restricted Hartree-Fock,\cite{lee_twenty_2022,awasthi2025noncovalent} coupled cluster wavefunctions projected onto CI subspaces,\cite{mahajan_beyond_2025,kjonstad2025systematic} and strategies involving single-shot selected CI calculations following active-space orbital optimization\cite{vuong2026evaluating} -- there is still a lack of consensus for strongly correlated electronic states in, e.g., polynuclear transition metal compounds.  
The myriad approaches of multiple active research groups to obtaining the trial might be classified into two classes.  The first relies on multi-determinant trial wavefunctions that approximate the exact wavefunction, including truncated full CI,\cite{huang2024gpu} selected CI trials,\cite{mahajan_selected_2022,malone_ipie_2023} and matrix product states.\cite{jiang_unbiasing_2025}  The second posits that spin symmetry-breaking at the mean-field level can provide a sufficiently appropriate trial wavefunction for certain classes of strongly correlated states.\cite{danilov2025capturing,qin2016benchmark} 
Indeed, from a computational cost perspective the latter approach is clearly preferable as the cost, in general, of the former trial wavefunctions scales exponentially with system size.
However, as will be explored in this work, a critical challenge of the second approach is the fact that Hartree-Fock (even in the simplest molecular models of strong correlation) has a complex solution landscape littered with myriad nearly-degenerate local minima.\cite{burton_energy_2021} 
ph-AFQMC can be highly sensitive to the choice of solution used as the trial wavefunction; for example, in a recent work we examined the dissociation of O$_2$  with ph-AFQMC-GHF\cite{danilov2025capturing} and identified multiple GHF solutions separated by $\mathcal{O}$(1) mHa that led to an unexpectedly wide range of ph-AFQMC energies.  The situation is much more severe in polynuclear transition metal compounds, \emph{vide infra}. 

The motivating question that the present work endeavors to address is:  of the combinatorially-many mean-field solutions that exist in strongly correlated states, how can one choose the most optimal unrestricted Hartree-Fock wavefunction as a trial for ph-AFQMC or as a reference state for low-order coupled cluster models?  Our data suggests that the answer is \emph{not} necessarily the single determinant with the lowest variational energy, although many times this is the case.  Corroborating  prior work focused on molecular non-valence correlation-bound anions,\cite{upadhyay2020role} we find that quantities reflecting the chemically or physically sensible distribution of electron density are more useful descriptors -- in this work we use  atomic spin densities (or the dipole moment for the linear He$_2^+$ model) to pinpoint the trial or reference state that best reflects either rigorous physical symmetries or oxidation state assignments / spin coupling patterns supported by experimental data.  In both ph-AFQMC and CCSD(T) this leads to the most accurate ground-state energies vs.\@ either exact FCI energies or theoretical best estimates.

The non-monotonic energy trajectory that eventually lands on an overly-negative ph-AFQMC energy observed for the [4Fe--4S]$^{4+}$ state motivated us to investigate the evolution of walkers under the phaseless constraint. 
Our analysis indicates that, in analogy with the stretched He$_2^+$ case, the stochastic wavefunctions in all three Fe-S systems investigated exhibit an overlap with the trial wavefunction that quickly decays to zero, which suggests suboptimal guidance of the walker trajectories in imaginary time by the trial state (importance sampling) and that local energies are, on average, spuriously negative. This explains the, at times, overly negative ph-AFQMC energies, and the contemporaneous finding of Chan and coworkers that ph-AFQMC energies for Fe-S systems can become more positive when correlated trial wavefunctions are used\cite{eirik_f_kjonstad_can_2026}.  In our view, the detailed elucidation of how ph-AFQMC with certain single-determinant UHF trial wavefunctions produces such accurate energies opens up potential avenues for future investigation of trial wavefunction forms and alternative/additional constraints.

\section{Methods}

\subsection{The phaseless constraint}

Imaginary-time propagation of an initial state can be exactly mapped onto an open-ended random walk in non-orthogonal Slater determinant space.\cite{zhang1995constrained,zhang2003quantum,motta_ab_2018,lee_twenty_2022} 
The Monte Carlo wavefunction $\ket{\Psi_{\text{MC}}(\tau)}$, defined at a point in imaginary-time $\tau$, can be expressed as a linear combination of random walkers, each (indexed $i$) with a weight $w_i(\tau)$ and a determinant $\ket{\varphi_i(\tau)}$:
\begin{equation}
\ket{\Psi_{\text{MC}}(\tau)} \approx \sum_i w_i(\tau)\frac{\ket{\varphi_i(\tau)}}{\braket{\psi_T | \varphi_i(\tau)}}.
\label{eqn:qmc_wfn}
\end{equation}

The integral propagator $e^{-\tau\hat{H}}$ can be modified with an importance sampling transformation, where the integration variable (auxiliary-field, $\mathbf{x}$) is shifted by a so-called force bias, $\bar{\mathbf{x}}$, chosen such that the phase  of the overlap ratio, $\frac{\langle \psi_T | \hat{B}(\mathbf{x}, \mathbf{\bar{x}}) | \varphi_i \rangle}{\langle \psi_T | \varphi_i \rangle}$, vanishes up to $\mathcal{O}(\Delta\tau)$.  The weight update factor is:  
\begin{equation}
    \frac{\langle \psi_T | \varphi_i(\tau) \rangle}{\langle \psi_T | \varphi_i(\tau-\Delta\tau) \rangle} * e^{\frac{1}{2}\bar{\mathbf{x}}^2 - \mathbf{x}\cdot \bar{\mathbf{x}}} \approx e^{-\Delta\tau E_L}
    \label{eqn:weightupdate}
\end{equation} 
where the local energy is $E_L(\tau) = \frac{\langle \psi_T|\hat{H}|\varphi_i(\tau)\rangle}{\langle \psi_T|\varphi_i(\tau)\rangle}$.  The ``hybrid'' formalism uses the left-hand side of Equation \eqref{eqn:weightupdate} explicitly, and the specific choice of the force-bias ensures that the phases of these two factors cancel to leading order.\cite{shi_symmetry_2013}  

However, even when using small imaginary-time steps and the weight update in Equation \eqref{eqn:weightupdate}, MC estimates will still inevitably suffer from an exponentially vanishing signal-to-noise ratio.  
To understand this, consider the MC estimate of the energy: 
\begin{equation}
    E_{\text{MC}}(\tau) = \frac{\sum_i w_i(\tau)E_{\text{L}}^i(\tau)}{\sum_i w_i(\tau)} 
\end{equation}
Each walker weight is a stochastically fluctuating complex number that acquires an arbitrary phase.  For systems where the propagator and thus walker weights and orbitals are real-valued, the phase of each walker weight can be only 0 or $\pi$, leading to weights being either along the positive or negative real axis.  When $E_{\text{MC}}$ is calculated the contributions from walkers with positive and negative weights will tend to cancel, resulting in a loss of MC signal.  The constrained-path approximation keeps only walkers with a positive overlap with a fixed trial wavefunction.\cite{zhang1995constrained}  Analogously, but more generally, in the case of phaseless AFQMC, the phaseless constraint involves taking the norm of the weight update factor (or real part of the local energy) in Equation \eqref{eqn:weightupdate} and then multiplying this by $\max\big( 0, \cos(\Delta\theta)\big)$, where $\Delta\theta = \arg \left[ \frac{\braket{\psi_T | \varphi_i(\tau)}}{\braket{\psi_T | \varphi_i(\tau - \Delta \tau)}}\right]$.  The phaseless constraint, which relies on optimal choice of the force bias to keep (unconstrained) phase rotations of the weight update factor small, keeps track of the overlap of the walker determinants with a fixed trial wavefunction to break the rotational invariance in the complex plane (i.e., $\varphi$ and $e^{i\gamma}\varphi$ are physically identical).  It defines a gauge choice ($\gamma = 0$) such that   the $\{w_i\}$ in Equation \eqref{eqn:qmc_wfn} are constrained to be real and positive, and walker determinants that accumulate a phase in excess of $\pm \pi/2$ are removed.  In the zero time step limit, the phaseless formalism was designed to prevent a finite population at the origin of the complex plane defined by $\langle\psi_T|\varphi\rangle$, but in practice this cannot be strictly enforced.  

When the trial wavefunction is an exact eigenstate of the Hamiltonian, the phaseless constraint does not bias the ph-AFQMC energies.  For weakly correlated systems, as the trial approaches the limit of exactness the phaseless bias can be reduced systematically; however, in the regime of strong correlation this may not always be the case.  One may wonder which property of the trial wavefunction must systematically improve to reduce the bias in ph-AFQMC energies.  In this work we would like to shed light on how to choose an optimal trial wavefunction for ph-AFQMC among the class of unrestricted Hartree-Fock states.  For the Fe-S clusters investigated we find that a judicious choice of UHF trial enables ph-AFQMC energies to be very close to the reference values, despite a rapid decay in the overlap between the stochastic MC wavefunction and the trial.

\subsection{Fe-S models}

Extensive experimental efforts have been devoted to the synthesis and characterization of Fe-S cluster models\cite{beinert_iron-sulfur_1997,skeel2025iron} that mimic the 
chemical and physical motifs most commonly found in biological systems.\cite{johnson_structure_2005}
Early demonstrations of synthetic Fe-S clusters in the 1970s\cite{Holm1977} have been followed by a 
wide array of different model systems.\cite{Rao2004}
Multiple spectroscopic techniques including electron paramagnetic resonance (EPR),\cite{Palmer1966} Mossbauer,\cite{Blomstrom1964} and nuclear magnetic resonance (NMR)\cite{Poe1970} 
have been essential for deducing the structure and properties 
of these clusters since the earliest investigations of Fe-S proteins. Beyond experiment, phenomenological models of the [2Fe--2S] \cite{noodleman1984electronic, noodleman1995orbital, noodleman1986ligand, Papaefthymiou1982} and [4Fe--4S] \cite{papaefthymiou1987moessbauer, blondin1990interplay} clusters have likewise been used to explore the electronic and magnetic character of these states. The history and limitations of these models have been well reviewed in a more recent \textit{ab initio} investigation of the complexes \cite{sharma_low-energy_2014}, a work which corroborates the prediction that the iron centers of the [2Fe--2S]$^{2+}$ cluster are antiferromagnetically coupled.\cite{gibson1966iron, brintzinger1966ligand,dunham1971structure,bertini1991proton}.

For the [4Fe--4S]$^{2+}$ cluster, while simple electron counting would suggest the presence of two Fe(III) ions and two Fe(II) ions, spectroscopic experiments find that each Fe oxidation state is effectively 2.5. \cite{Holm1974}
Structural studies have established that the cubane is distorted along one axis, and the pair of Fe atoms on the top or bottom surface of the cube is ferromagnetically coupled into an S=$\frac{9}{2}$ state, as indicated in Figure \ref{fig:4fe4s_summary}.  The intra-pair ferromagnetic coupling has been explained by a ``double exchange'' mechanism,\cite{Zener1951,Anderson1955,Girerd1989,Noodleman1991a,Noodleman1991b} and the two formally mixed-valence irons in a ferromagnetic pair are experimentally found to be indistinguishable, implying even charge delocalization between the two-centers.  In turn, the two ferromagnetic pairs (one on the top face, the other on the bottom face) are antiferromagnetically coupled resulting in an overall spin singlet ground-state.\cite{Papaefthymiou1982,beinert_iron-sulfur_1997,sharma_low-energy_2014}

The third system we study is the highly oxidized [4Fe--4S]$^{4+}$ state, with four antiferromagnetically coupled Fe(III) centers.  While irrelevant to biological Fe-S processes, it has been accessed experimentally in a synthetic cluster\cite{ohki_synthetic_2011} and its properties have been explored theoretically\cite{grunwald_vibrational_2025,lee_evaluating_2023}. 

We have used the 30e20o [2Fe--2S]$^{2+}$ and 54e36o [4Fe--4S]$^{2+}$ active space models (based on localized high-spin DFT orbitals) 
as provided by Li and Chan\cite{li_spin-projected_2017}. We used the DMRG reference for [2Fe--2S]$^{2+}$ from the same report, however for [4Fe--4S]$^{2+}$ a more recent work by Zhai et al.\cite{zhai2026classical} has suggested  theoretical best estimates for three low-lying spin isomers based on extrapolated unrestricted DMRG and relatively high-order coupled cluster calculations. For the [4Fe--4S]$^{4+}$ state we used the 52e36o model and converged DMRG reference energy from Lee et al.\cite{lee_evaluating_2023}.

\begin{figure}[H]
    \centering
    \includegraphics[width=0.95\linewidth]{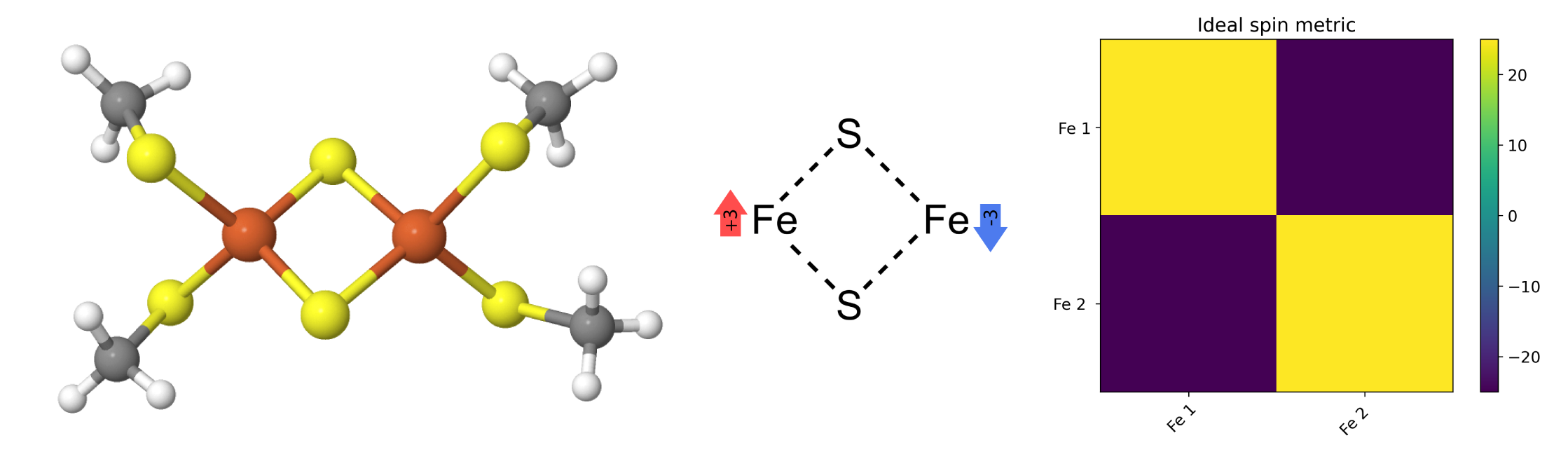}
    \caption{\footnotesize Cluster geometry (left), idealized spin configuration (middle), and corresponding reference spin density metric (right) for [2Fe--2S]$^{2+}$.}
    \label{fig:2fe2s_summary}
\end{figure}

\begin{figure}[H]
    \centering
    \includegraphics[width=0.95\linewidth]{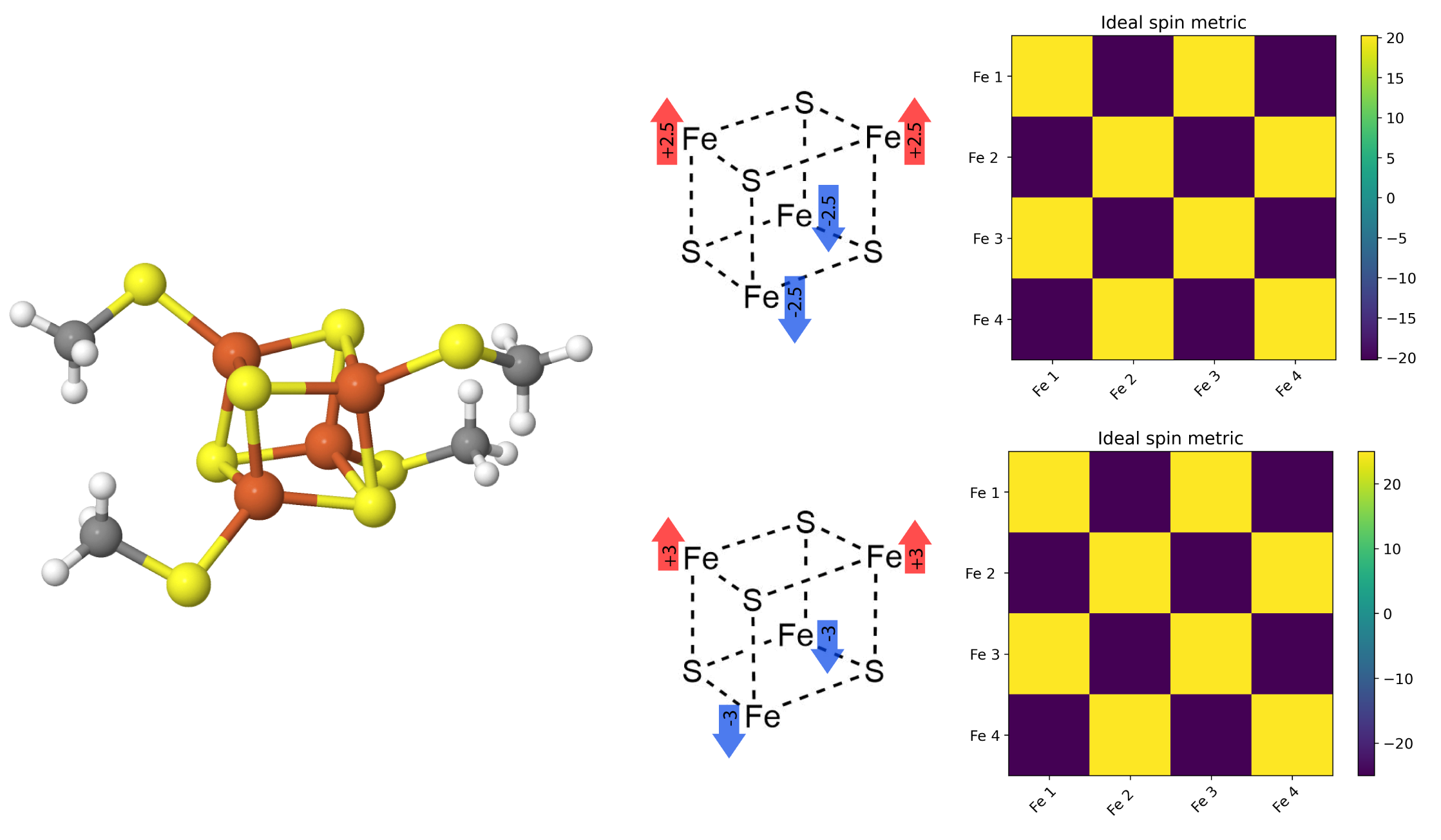}
    \caption{\footnotesize Same as Figure \ref{fig:2fe2s_summary} but for [4Fe--4S]$^{2+}$ (top) and [4Fe--4S]$^{4+}$ (bottom) states.}
    \label{fig:4fe4s_summary}
\end{figure}

\subsection{Computational Details}
To locate the various stable UHF solutions (that were later used as references in coupled cluster and trials in ph-AFQMC) we used PySCF's\cite{sun2018pyscf} Newton solver and stability analysis. We initialized our UHF solver with a one-electron reduced density matrix guess based on one of the following two strategies:
\begin{itemize}
    \item Manual specification of the orbital occupancies (generated by enumerating all the different possible combinations of electrons on the iron atoms)
    \item Randomized guess generated by performing 10 random Givens rotations (each with an equal probability to occur either in the $\alpha$ or $\beta$ sector of the density) between random orbitals with an angle chosen by drawing from a normal distribution $N(\pi/2, 1)$
\end{itemize}
Note that in our solution searches we discarded isoenergetic solutions (within 1$\mu$Ha) and limited correlated calculations to a low-energy subset.

To further ``fingerprint'' the UHF solutions we consider the net spin, magnitude, and orientation of the spin density operator:
\begin{equation}
\hat{\boldsymbol{s}}(\mathbf{r}) = \frac{1}{2} \sum_{\sigma \sigma'} \hat{\psi}^\dagger_\sigma(\mathbf{r}) \, \boldsymbol{\sigma}_{\sigma \sigma'} \, \hat{\psi}_{\sigma'}(\mathbf{r})
\end{equation}
The components of the expectation value of this vector may be obtained by contraction of the density
\[
\rho(\mathbf{r}) =
\begin{pmatrix}
\langle \hat{\psi}_\alpha^\dagger(\mathbf{r}) \hat{\psi}_\alpha(\mathbf{r}) \rangle & \langle \hat{\psi}_\beta^\dagger(\mathbf{r}) \hat{\psi}_\alpha(\mathbf{r}) \rangle \\
\langle \hat{\psi}_\alpha^\dagger(\mathbf{r}) \hat{\psi}_\beta(\mathbf{r}) \rangle & \langle \hat{\psi}_\beta^\dagger(\mathbf{r}) \hat{\psi}_\beta(\mathbf{r}) \rangle
\end{pmatrix} = 
\begin{pmatrix}
\mathbf{P}_{\alpha\alpha} & \mathbf{P}_{\alpha\beta} \\
\mathbf{P}_{\beta\alpha} & \mathbf{P}_{\beta\beta}
\end{pmatrix}
\]
with the components of the Pauli matrices $\boldsymbol{\sigma}_{\sigma \sigma'}$:
\begin{equation}
\sigma_x = 
\begin{pmatrix}
0 & 1 \\
1 & 0
\end{pmatrix}
, 
\sigma_y = 
\begin{pmatrix}
0 & -i \\
i & 0
\end{pmatrix}
, 
\sigma_z = 
\begin{pmatrix}
1 & 0 \\
0 & -1
\end{pmatrix}
\end{equation}
The symmetry restrictions in ROHF and UHF cases ensure that the off-diagonal elements are  $\mathbf{P}_{\alpha\beta}=\mathbf{P}_{\beta\alpha}^{\dagger}=0$, and formally only the $S_z$ contribution of the spin density vector has a surviving term through contraction with $\sigma_z$:
\begin{equation}
\hat{\boldsymbol{s}}(\mathbf{r})_{z} =
 \text{Tr}\left[\begin{pmatrix}
\mathbf{P}_{\alpha\alpha} & 0 \\
0 & \mathbf{P}_{\beta\beta}
\end{pmatrix}
\begin{pmatrix}
1 & 0 \\
0 & -1
\end{pmatrix}\right]
= \mathbf{P}_{\alpha\alpha} - \mathbf{P}_{\beta\beta}
\end{equation}
As the $S_x$ and $S_y$ components are exactly zero for these cases, this immediately determines the total net spin and the unidirectional orientation/magnitude of the collinear case for each UHF trial determinant under study.

The number of electrons of a specific spin may be obtained by the partial trace over the orbitals $\mu$ belonging to atomic center $A$, such that the Mulliken spin density of the center is determined by 
\begin{equation}
    \rho_s(A) = \sum_{\mu \in A} (\mathbf{P}_{\alpha\alpha} - \mathbf{P}_{\beta\beta})_{\mu\mu} S_{\mu\mu}
\end{equation}
where we have contracted the surviving elements of the spin density vector $\hat{\boldsymbol{s}}(\mathbf{r})_{z}$ with the (often orthonormal) elements of the overlap matrix $\boldsymbol{S}$ associated with $A$. The Mulliken spin density has historically been related to the gross electron population and the free valency on atomic centers\cite{mayer1983towards,mayer1984bond} and has seen demonstrable success in characterizing oxidation states\cite{blomberg1997comparative,kubin2018probing}.

To describe the deviation from a reference spin density pattern, we introduce a spin density metric (SDM) defined as the Frobenius norm of the deviation of the outer product of the UHF Mulliken spin densities from that of the ideal (denoted $I$ in the equation below) spin densities shown in the middle panels of Figures \ref{fig:2fe2s_summary} and \ref{fig:4fe4s_summary}: 

\begin{equation}
    \text{SDM} = \left|\left|\frac{(\rho \otimes \rho) - (\rho_I\otimes \rho_I)}{\rho_I\otimes \rho_I}\right|\right|_F.
\end{equation}

CCSD and CCSD(T) calculations were performed with PySCF\cite{sun2018pyscf}. 
While initial ph-AFQMC studies of the He$_2^+$ cation (specifically the calculations presented in Figure \ref{fig:He2p}) used the ipie program,\cite{malone_ipie_2023,jiang2024improved} the rest used an in-house code.  ph-AFQMC calculations used an imaginary-time step of $\Delta\tau = 0.005$ and a Cholesky decomposition threshold of $10^{-12}$.  We initialized the walkers with the same UHF state used as trial wavefunction throughout. For [2Fe--2S]$^{2+}$ and [4Fe--4S]$^{2+}$ autocorrelation effects were removed by a reblocking procedure where the optimal equilibration time was chosen such that the lowest ph-AFQMC energy is obtained.  For [4Fe--4S]$^{4+}$, where in some cases the trajectories were non-monotonic and relatively noisy, we report energies from reblocking analyses targeting the lowest statistical error.

To quantitatively assess the stochastically projected many-body wavefunction, $|\Psi_{\text{MC}}\rangle$, we compute the normalized squared overlap, $|\langle \bar{\psi}_T | \bar{\Psi}_{MC} \rangle|^2$, 
with the trial wavefunction (or another wavefunction of interest) using the following equation:
\begin{equation}
|\langle \bar{\psi}_T | \bar{\Psi}_{MC} \rangle|^2 = \frac{|\langle \psi_T | \Psi_{MC} \rangle|^2}{\langle \psi_T | \psi_T \rangle \langle \Psi_{MC} | \Psi_{MC} \rangle}.
\end{equation}
A population of 1024 walkers was used for calculating overlaps on He$_2^+$, while 2048 were used for the Fe-S clusters with the walkers written out every other block of 20 imaginary-time steps. 

\section{Results and Discussion}

\subsection{[2Fe--2S]$^{2+}$}

\begin{figure}[H]
    \centering
    \includegraphics[width=0.75\linewidth]{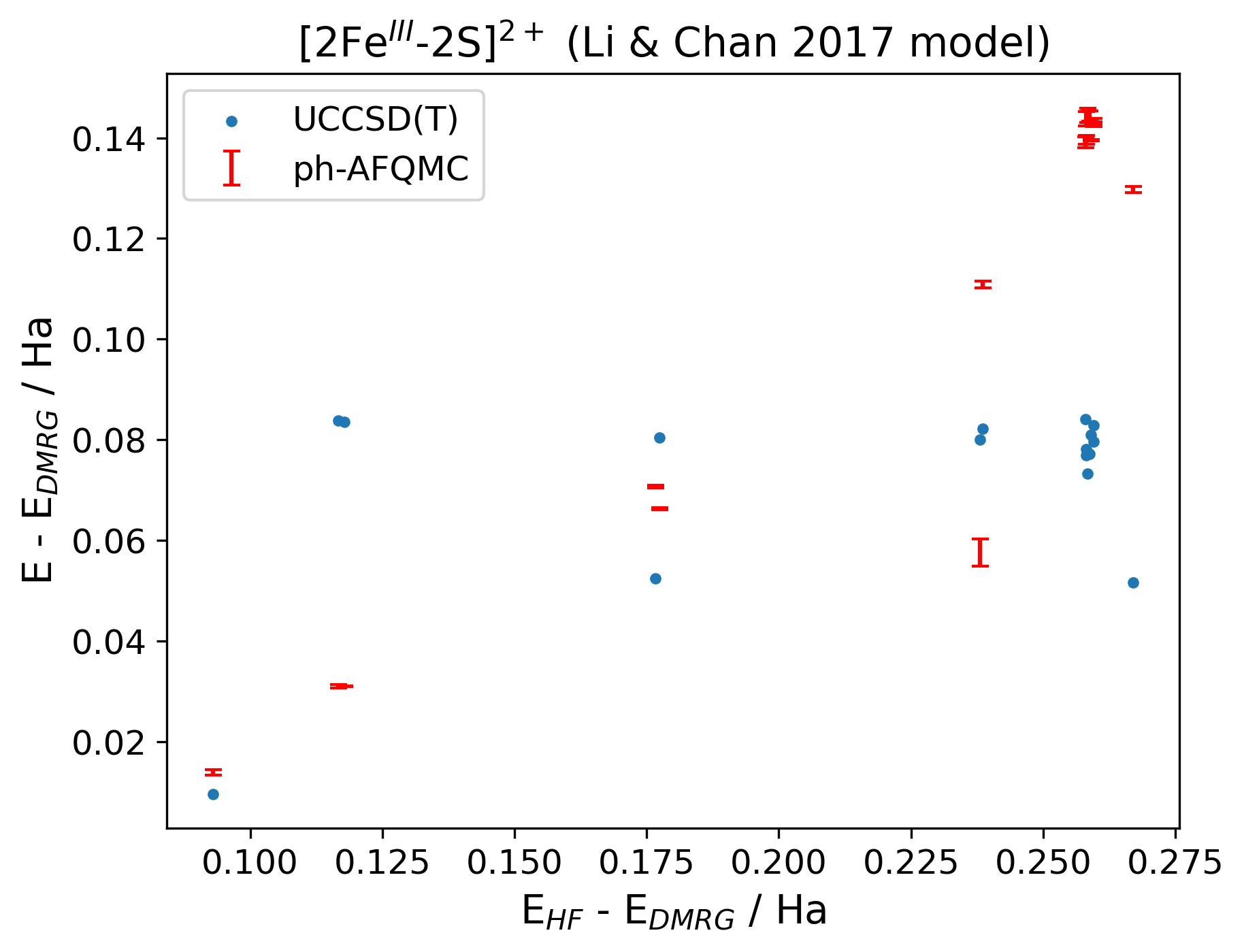}
    \caption{\footnotesize Deviations of the ph-AFQMC-UHF and UCCSD(T) energies from the DMRG reference, as a function of the deviation of the UHF trial/reference energy from DMRG.  
    }
    \label{fig:2FeEnergy}
\end{figure}

We begin by investigating the low-lying UHF states of the diferric complex, [2Fe(III)--2S]$^{2+}$. Figure \ref{fig:2FeEnergy} shows the deviations of the resulting ph-AFQMC and CCSD(T) energies from the DMRG reference value, plotted against the energy error of the UHF trial/reference. The ph-AFQMC energy when using the lowest-energy UHF state deviates from the reference value by $13.8\pm0.6$ mHa; this result from a single determinant trial wavefunction is encouraging and arguably unexpected, given previous studies that required on the order of a million determinants to achieve a similar level of accuracy\cite{huang2024gpu}. UCCSD(T) is slightly more accurate with this UHF reference state, above DMRG by 9.6 mHa, while the UCCSD energy deteriorates to a 25.1 mHa error. Figure \ref{fig:2FeEnergy} shows that UCCSD(T) is relatively insensitive to the reference used, with most CCSD(T) energies clustering between 52-84 mHa above DMRG despite more significant differences in the underlying UHF energies. That the UCCSD(T) energy using the lowest-energy HF reference is singularly outside of this energy range, and is substantially improved with respect to the DMRG reference value, goes against the common wisdom that CC is ``orbital insensitive.'' \cite{bartlett2007coupled}  

The experimentally-deduced oxidation states and spin coupling pattern, along with the corresponding reference SDM, of the [2Fe--2S]$^{2+}$ cluster are shown in Figure \ref{fig:2fe2s_summary}.  Figure S1 confirms the trend that lower energy UHF states correspond to more appropriate iron spin densities vs.\@ the idealized distribution.   
In this specific case the lowest-energy UHF trial wavefunction is the state with the smallest SDM value, which leads to the most accurate ph-AFQMC energy.

\subsection{[4Fe--4S]$^{2+}$}

\begin{figure}[H]
    \centering
    \includegraphics[width=0.85\linewidth]{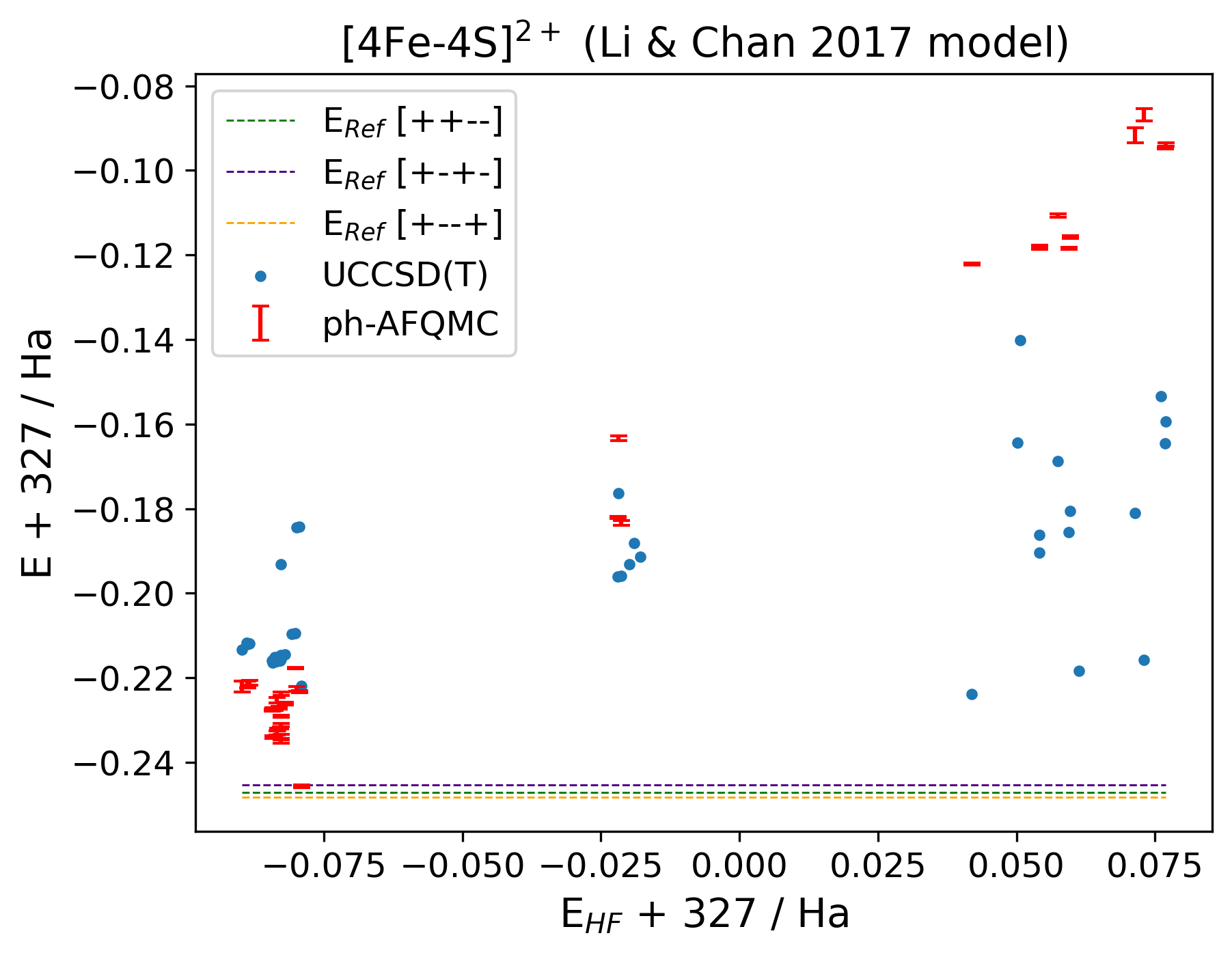}
    \caption{\footnotesize ph-AFQMC-UHF and UCCSD(T) energies as a function of the UHF trial/reference energy. The three reference theoretical best estimates are $E_{\text{Ref}}$[++-- --] $=-327.247062\pm 0.000984$, $E_{\text{Ref}}$[+--+--]$=-327.245289\pm0.000733$, $E_{\text{Ref}}$[+-- --+]$=-327.248199\pm0.001820$ from Zhai et al \cite{zhai2026classical}, wherein the 3 solutions are referred to as BS1, BS2, and BS3 respectively (the error bars are omitted for clarity).} 
    \label{fig:4FeEnergy}
\end{figure}

We now consider the mixed-valence [4Fe--4S]$^{2+}$ cubane system, which was recently shown to have three low-lying spin isomers separated by at most a few mHa.\cite{zhai2026classical}  Figure \ref{fig:4FeEnergy} shows the ph-AFQMC and CCSD(T) energies corresponding to the different UHF states found, along with the theoretical best estimates for the three spin isomers. The UCCSD(T) energies form roughly a horizontal band ranging from 20 to 100 mHa above the references.  In contrast, a nearly linear behavior can be seen in the case of the ph-AFQMC energies, as was found also for [2Fe-2S]$^{2+}$ above.  Notably, there is one UHF state that leads to remarkable agreement with the theoretical best estimates, which is \emph{not} the lowest-energy UHF solution.
The fact that there is a UHF trial that leads to a near-exact ph-AFQMC energy is at once extraordinary and vexing, in light of previous studies that illustrated the difficulty of converging the energy of the smaller [2Fe--2S]$^{2+}$ system using multi-determinant trial wavefunctions.\cite{huang2024gpu,danilov_enhancing_2025} While this result will be analyzed further in Section \ref{He2psection}, we now look closer at how this optimal UHF trial might be identified among the myriad possible UHF states.

\begin{figure}[H]
    \centering
    \includegraphics[width=0.75\linewidth]{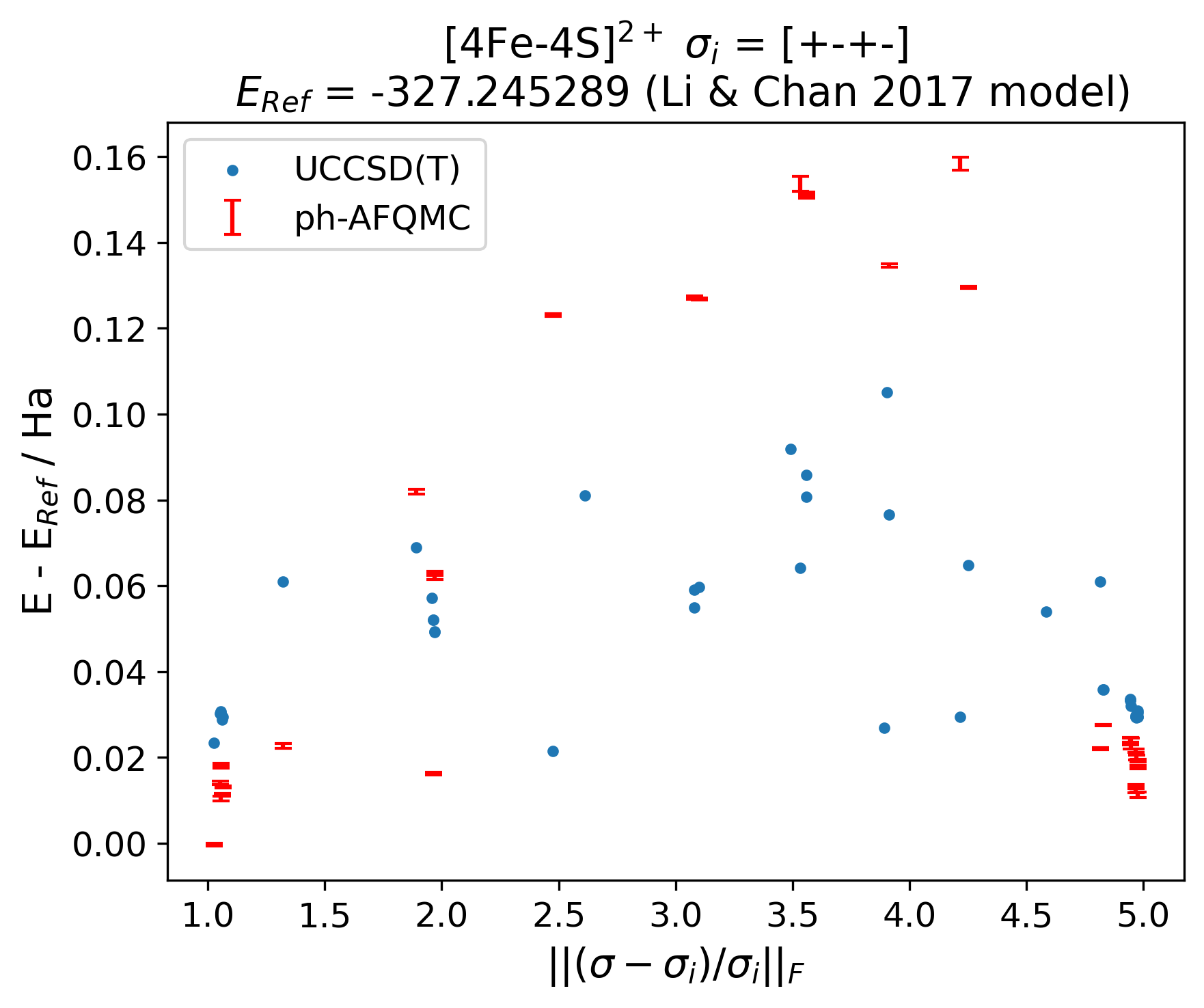}
    \caption{\footnotesize Deviations of the ph-AFQMC-UHF and UCCSD(T) energies from the theoretical best estimate\cite{zhai2026classical} of the $[+-+-]$ spin isomer, as a function of the SDM of the underlying UHF state.  We consider the ideal spin distribution, $\sigma_I$, to be Fe$_1^{\uparrow 2.5}$ Fe$_2^{\downarrow 2.5}$ Fe$_3^{\uparrow 2.5}$ Fe$_4^{\downarrow 2.5}$.}
    \label{fig:4FeSDM}
\end{figure}

Focusing on what we will denote the $[+-+-]$ spin isomer (which corresponds to the spin configuration shown in Fig.  \ref{fig:4fe4s_summary}), Figure \ref{fig:4FeSDM} plots the ph-AFQMC and CCSD(T) energy errors against the SDM of the UHF trial/reference wavefunctions, using the $[+-+-]$  spin-density pattern as ``ideal''.  
In the SDM range of 1-4.5 the trend in the ph-AFQMC energy error grows as the SDM increases.  Moreover, it is clear that the UHF state with the lowest SDM is the best trial wavefunction, yielding a ph-AFQMC energy that agrees closely with the theoretical best estimate (and a well-behaved energy trajectory in imaginary-time).   
This optimal-SDM UHF trial is the 23rd excited UHF state among the UHF solutions that we found.  
Its atomic spin-densities are +3.9, -3.9, +3.9, -3.9 (where indices 1 and 3 correspond to the top face of the elongated cube, and indices 2 and 4 correspond to the bottom face).  
In a tetrahedral ligand field, Fe(II) and Fe(III) ions would be expected to correspond to 4 and 5 unpaired electrons, respectively; therefore the experimentally established 2.5 oxidation state would correspond to atomic spin density magnitudes of 4.5.  Although the atomic spin density magnitudes are slightly smaller than 4.5, the relative sign structure matches the targeted $[+-+-]$ spin isomer and the magnitudes on all four Fe ions are equal.

\begin{figure}[H]
    \centering
    \includegraphics[width=0.75\linewidth]{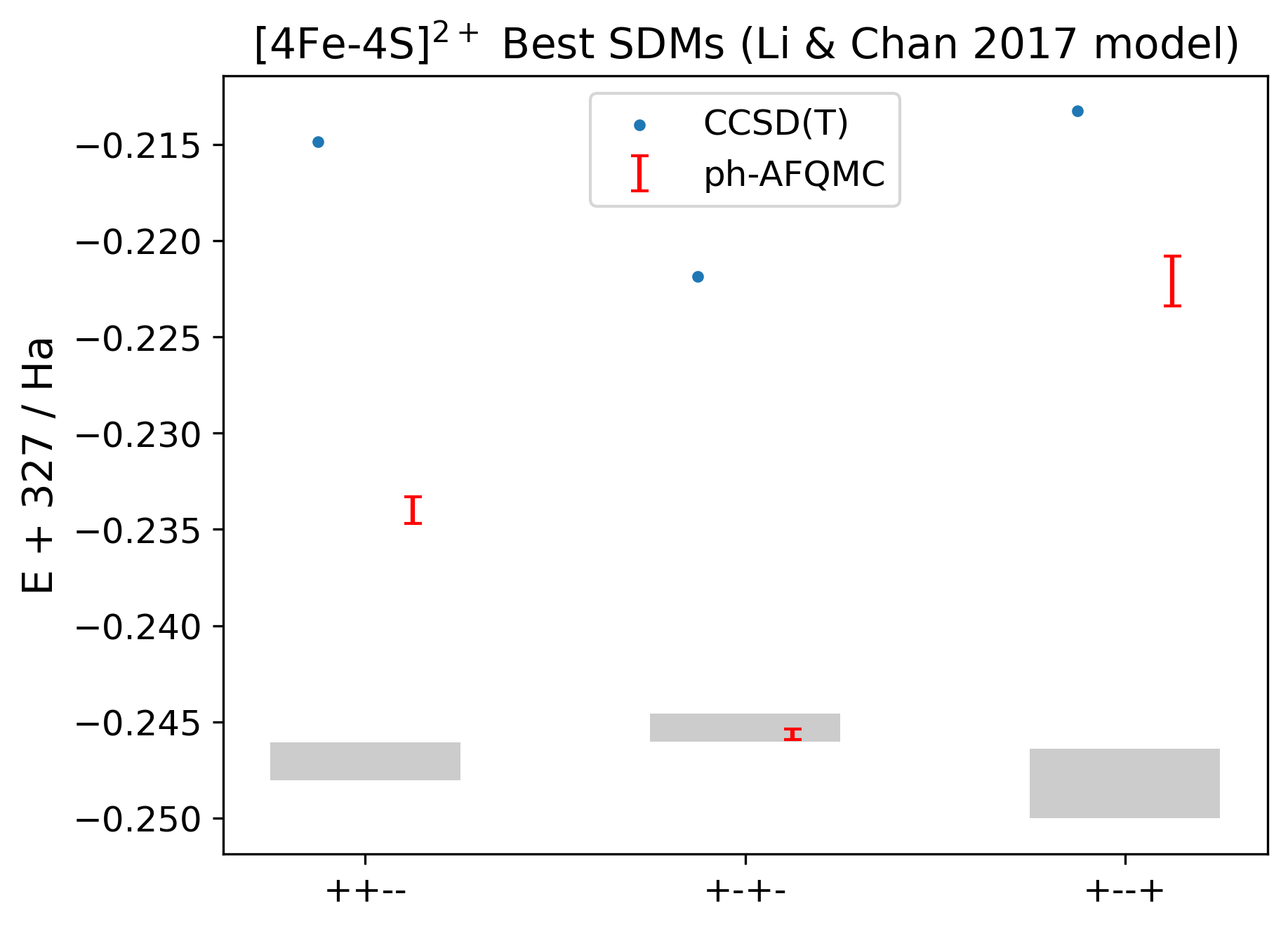}
    \caption{\footnotesize  ph-AFQMC and UCCSD(T) energies for the three spin isomers, using the UHF trial/reference state with the lowest spin-density metric. The grey boxes indicate the theoretical best estimates from Ref. \citenum{zhai2026classical}:  the top of the box represents the upper-bound derived from a UCC procedure, the bottom represents the lowest of the extrapolated UCC or UDMRG procedures from the same work.}
    \label{fig:4Fe-bestsdm}
\end{figure}

There is a cluster of UHF solutions that leads to competitively low ph-AFQMC energies in the SDM range 4.5-5.0.  These do not show ferromagnetically coupled pairs in the top and bottom cube faces antiferromagnetically coupled, but rather other spin coupling patterns involving the four Fe ions (as opposed to 
charge excitations from, e.g., the bridging sulfur atoms).  These are representative of other Hamiltonian eigenstates that are nearly degenerate with the ground-state, i.e. precisely the other two spin isomers that we will denote [$++--$] and [$+--+$].  
To target these states, 
we constructed ideal atomic spin densities corresponding to these two spin isomers, and select the UHF trial/reference with the lowest SDM.  Figure \ref{fig:4Fe-bestsdm} shows the resulting ph-AFQMC and CCSD(T) energies along with the theoretical best estimates for all three spin isomers.  We find that ph-AFQMC with the optimal-SDM UHF trials in the cases of [$++--$] and [$+--+$] leads to energies that are more positive than the reference values by $13.1\pm0.7$ and $26.1\pm1.3$ mHa, respectively.  Taken together, these results suggest that the near-perfect accuracy found for the [$+-+-$] spin isomer cannot be generally expected for all Fe-S systems using ph-AFQMC with optimal-SDM UHF trials.  Nevertheless, this ph-AFQMC protocol for the three [4Fe--4S]$^{2+}$ spin isomers clearly outperforms UCCSD(T).

\subsection{[4Fe--4S]$^{4+}$}
\begin{figure}[H]
    \centering
    \includegraphics[width=0.75\linewidth]{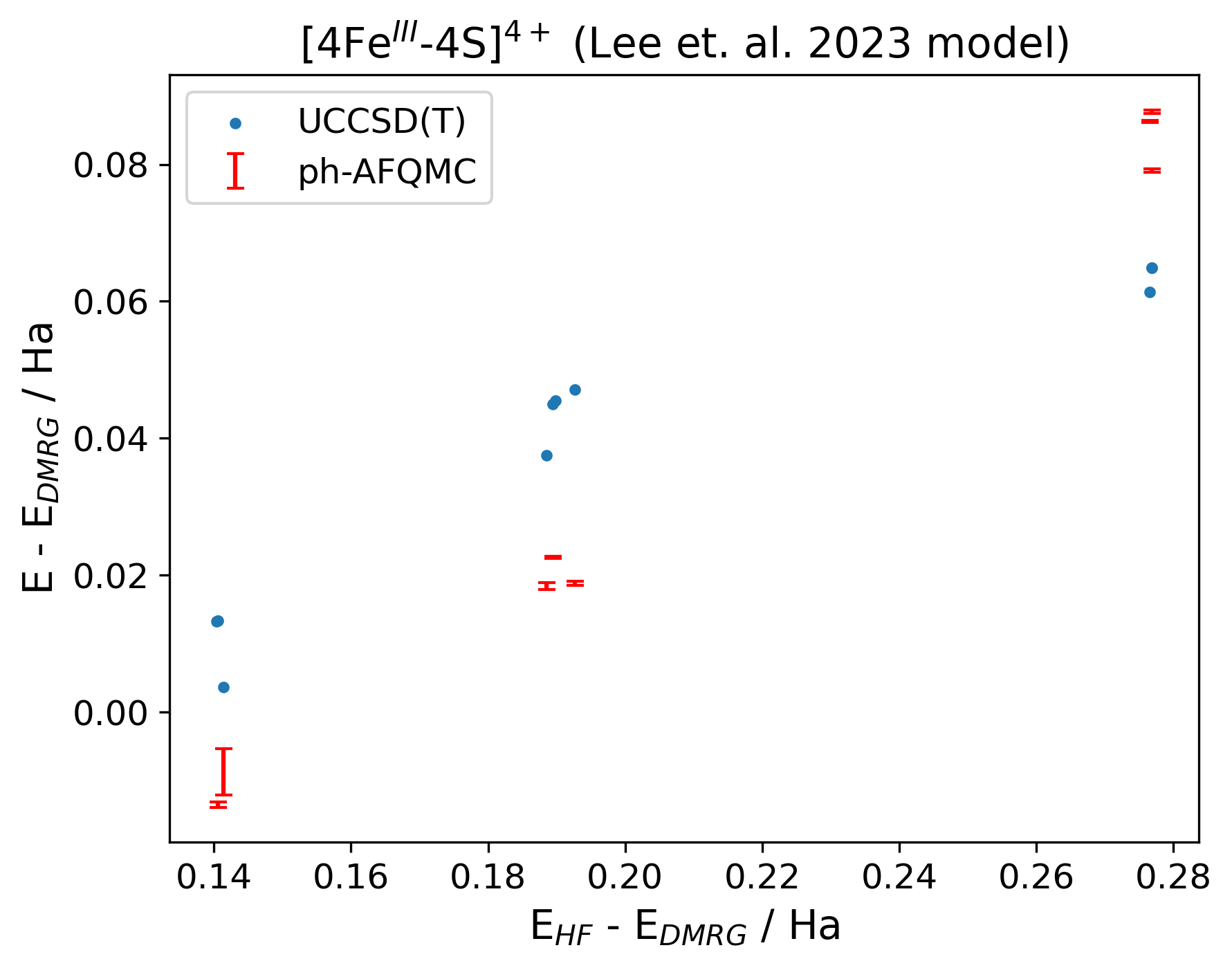}
    \caption{\footnotesize Deviations of the ph-AFQMC-UHF and UCCSD(T) energies from the DMRG reference, as a function of the deviation of the UHF trial/reference energy from DMRG. }
    \label{fig:4Fe3E}
\end{figure}

Turning now to the tetraferric [4Fe(III)--4S]$^{4+}$ state, Figure \ref{fig:4Fe3E} shows that the nearly linear trend of increasing ph-AFQMC energy error vs.\@ trial energy error is present here as well.  Notably, there is a UHF reference that allows UCCSD(T) to be very accurate, deviating from the DMRG reference by only 3.6 mHa.  The two lowest-energy UHF states, when used as trial wavefunctions in ph-AFQMC, result in energies that are too negative compared with the DMRG reference value.  

\begin{figure}[H]
    \centering
    \includegraphics[width=0.75\linewidth]{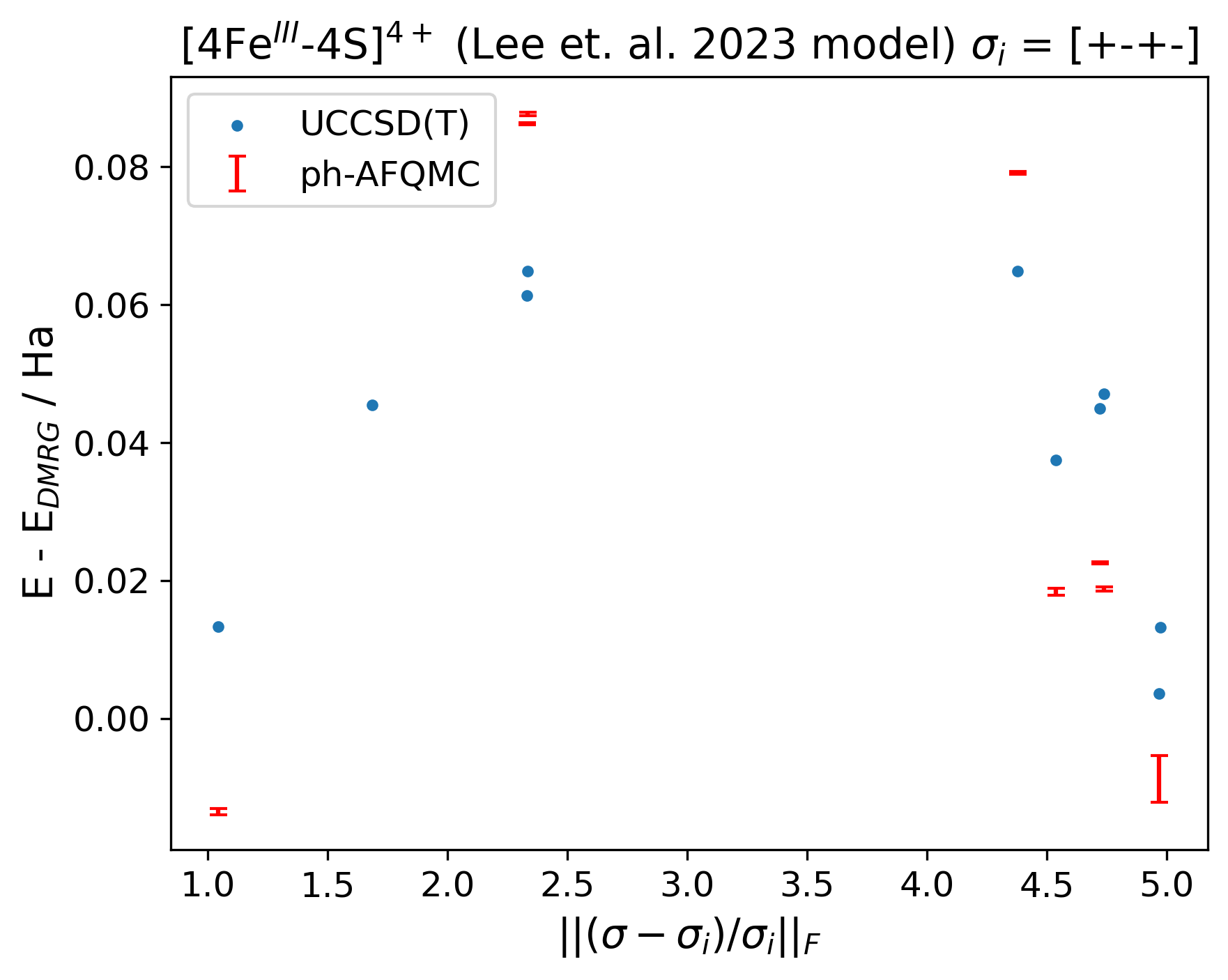}
    \caption{\footnotesize Deviations of the ph-AFQMC-UHF and UCCSD(T) energies from the DMRG reference, as a function of the SDM of the underlying UHF state. In this system we consider the ideal spin distribution, $\sigma_I$, to be Fe$_1^{\uparrow 3}$ Fe$_2^{\downarrow 3}$ Fe$_3^{\uparrow 3}$ Fe$_4^{\downarrow 3}$.} 
    \label{fig:4Fe3SDM}
\end{figure}

\begin{figure}[H]
    \centering
    \includegraphics[width=0.7\linewidth]{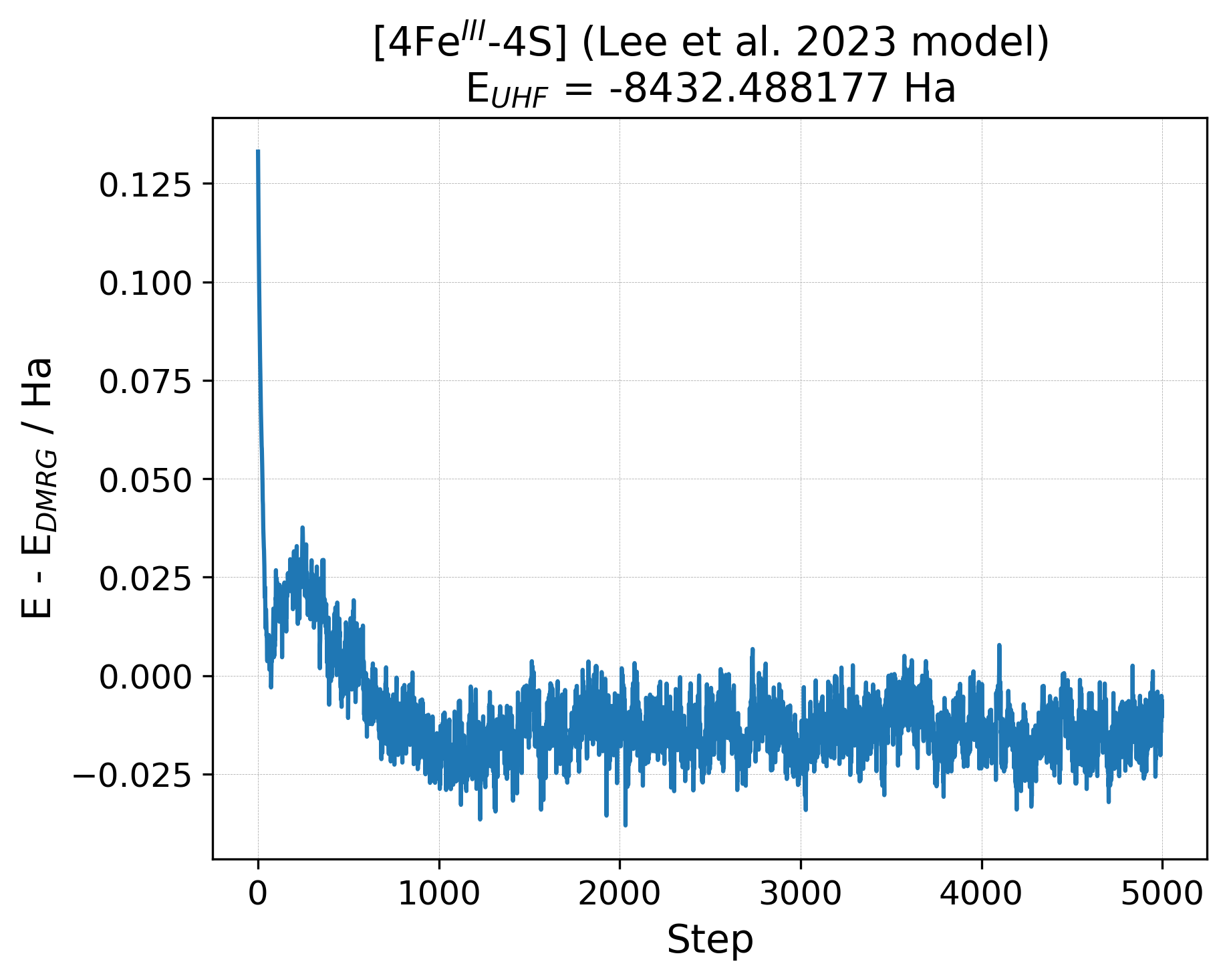}
    \caption{\footnotesize ph-AFQMC trajectory using the UHF trial with smallest SDM for the [4Fe--4S]$^{4+}$ state. Note the shallow dip early in the trajectory with a minimum near the reference DMRG value before a sharp rise and then slow decay to a relatively noisy plateau at an overly negative energy.}
    \label{fig:4Fe3badtraj}
\end{figure}

For this state we unfortunately do not have geometric or orbital information, as we simply performed calculations starting from an FCIDUMP file.  As a result there are three possible reference SDMs derived from the  [++-- --], [+--+--], and [+-- --+] orientations.   Figure \ref{fig:4Fe3SDM} takes the second as the reference SDM (choosing the other two leads to similar conclusions).  The two UHF trials with extreme values of SDM, i.e. around 1 and 5, lead to the most accurate ph-AFQMC energies:  8.8 $\pm$ 3 and 13.6 $\pm$ 0.4 mHa \emph{below} the DMRG energy, respectively.  Furthermore, the unusual energy vs.\@ imaginary-time trajectory when using the UHF trial with SDM $\sim$1 is shown in Figure \ref{fig:4Fe3badtraj}. 
The trajectory very quickly arrives at the exact energy before collapsing to a final plateau that is too low in energy. 
While ph-AFQMC is not variational, we will show later that an unexpected mechanism involving small local energy denominators is responsible for the overly negative ph-AFQMC energies (and the unexpected features in the energy trajectory in this case).  

The UHF state with the smallest SDM has atomic spin densities on the irons of magnitude 4.3.  The Fe(III) oxidation state implies 5 unpaired same-spin electrons per Fe ion, so here again the UHF spin density magnitudes are slightly smaller.  
The electron configuration represented by this trial state is only one of six possible (collinear) configurations that place (approximately) five aligned local spins on each of the four Fe centers while respecting the constraint of $S_z = 0$. Unlike the coupling patterns possible in the mixed-valence case, in which a ferromagnetic pair of Fe atoms is antiferromagnetically coupled to the other pair,  each ferric center of the [4Fe(III)--4S]$^{4+}$ complex interacts through a ``superexchange'' mechanism \cite{anderson1950antiferromagnetism, anderson1959new, kanamori1959superexchange, goodenough1968spin} to each of the others favoring antiferromagnetic coupling, resulting in geometric spin frustration wherein none of the local spins can be anti-aligned with all of its neighbors.  Despite the slight structural distortion of the cubane, it is well-established that such frustration gives rise to near-degeneracies of the many-body states, with six spin configurations dominating the low-energy manifold.
This implies that the overlap of the trial with the exact ground-state eigenfunction will be at most one-sixth (at most because other charge-excited/transferred configurations that contribute to the superexchange mechanism are not accounted for, and the UHF trial is strongly spin contaminated while the eigenstates are not).

\subsection{Analogy with stretched He$_2^+$}
\label{He2psection}

The dissociation of the helium dimer cation, He$_2^+$, is a  challenging model system for single-configurational electronic structure methods.  The lowest-energy UHF solution exhibits spontaneous symmetry breaking, artificially localizing the single unpaired electron onto one of the Helium atoms.  This UHF state has a dipole moment whereas the true wavefunction, which involves a linear combination of both possible localized solutions, does not.  With approximate density functionals such as PBE and PBE0,  self-interaction errors lead to a single-configurational state with zero dipole moment due to an unphysically delocalized electron/hole.\cite{mishra_study_2022, hait2018delocalization}

\begin{figure}[H]
    \centering
    \includegraphics[width=0.9\linewidth]{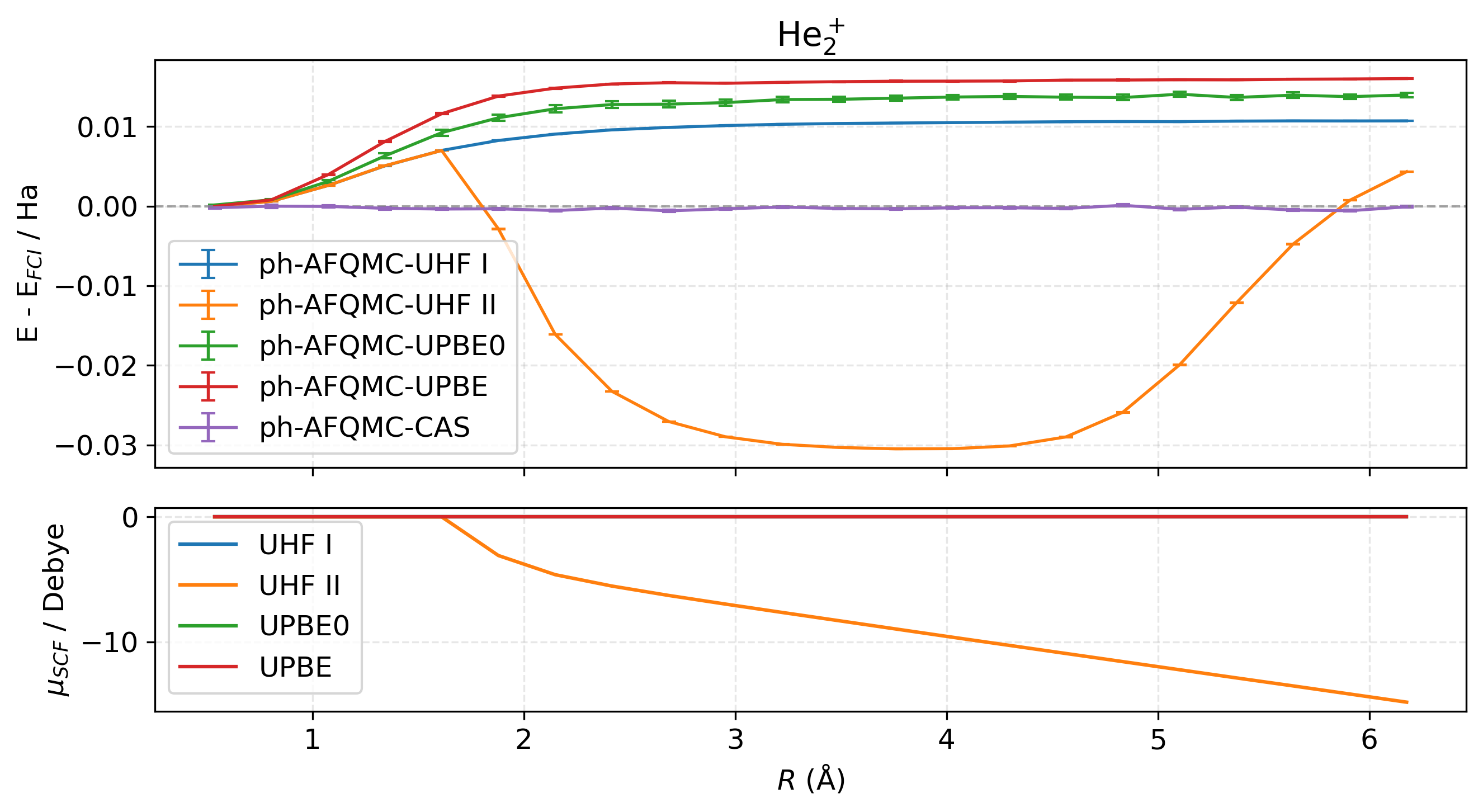}
    \caption{\footnotesize ph-AFQMC energies corresponding to various trial wavefunctions vs.\@ FCI along the dissociation coordinate of the helium dimer cation (top), and dipole moments of the single-determinant UHF and KS-DFT trials. (bottom) 
    }
    \label{fig:He2p}
\end{figure}

\begin{figure}[H]
    \begin{subfigure}[b]{0.6\textwidth}
        \centering
        \includegraphics[width=\textwidth]{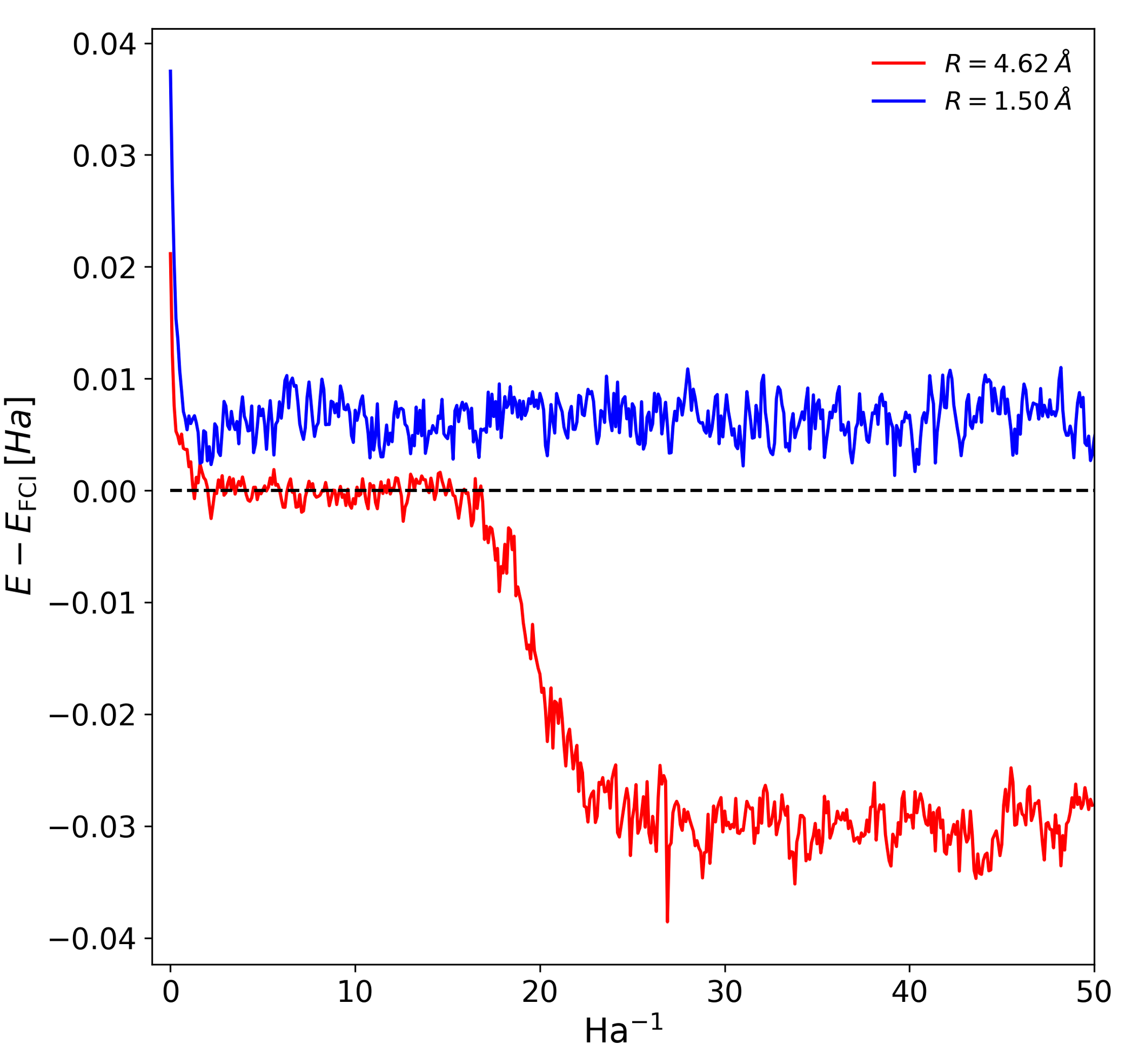}
      \caption{\footnotesize }
        \label{fig:sub1_he2_afqmc}
    \end{subfigure}
    \hfill
    \begin{subfigure}[b]{0.6\textwidth}
        \centering
        \includegraphics[width=\textwidth]{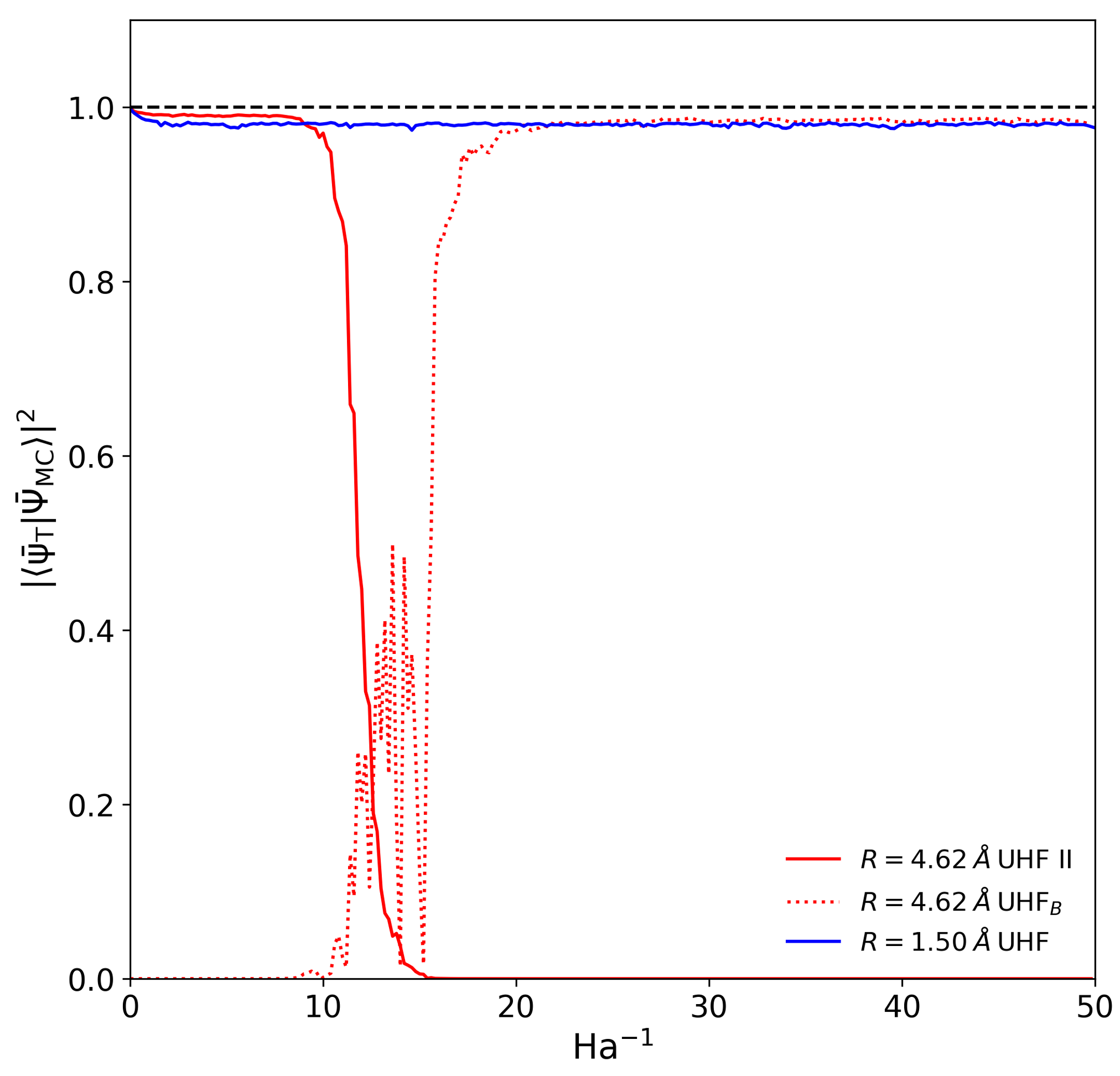}
        \caption{\footnotesize }
        \label{fig:sub2_he2_afqmc}
    \end{subfigure}
    \caption{\footnotesize (a) ph-AFQMC energy trajectories for the He$_2^+$ system at $R= 1.50 \text{\AA}$ and $R= 4.62 \text{\AA}$ with the UHF II trial wave function. Note the double-plateau structure for $R= 4.62 \text{\AA}$. (b) Normalized squared overlaps between the stochastic ph-AFQMC wavefunction and UHF states at the two bond lengths.}
    \label{fig:he2_afqmc_energy_overlaps}
\end{figure}

In this study we have located two low-lying UHF solutions which energetically separate at 1.75\AA \ and beyond.  The lowest energy state (denoted here as UHF II), stable to orbital rotations, has a growing dipole moment along the dissociation coordinate whereas the unstable higher-energy state (UHF I) retains a dipole moment of exactly zero. In actuality, UHF II is doubly-degenerate as one could localize the lone electron on either Helium atom.
Figure \ref{fig:He2p} shows the resulting ph-AFQMC dissociation curves, along with results using minimal (3e2o) CASSCF trials and also single determinant trials formed from occupied Kohn-Sham orbitals obtained with the PBE and PBE0 functionals.  While UHF I leads to under-correlated energies, which eventually plateau at $\sim$10 mHa above the exact value by 3\AA, UHF II (the trial state with the growing dipole) leads ph-AFQMC to over-correlate -- at worst by some 30 mHa just before 4\AA --  before returning toward the ph-AFQMC energies obtained from the UHF I trial.  The energy vs.\@ imaginary-time trajectory at a bond length of 4.62\AA \ is shown in Figure \ref{fig:he2_afqmc_energy_overlaps}; the ph-AFQMC energy quickly equilibrates to a transient plateau at the exact (FCI) energy, but then collapses to another lower-energy plateau that persists out to very long imaginary-times.  This is exactly the same behavior as what was found previously for the [4Fe--4S]$^{4+}$ case.  
Note that the energies shown in Figure \ref{fig:He2p} were obtained by averaging only the second energy plateau (if two were encountered).    

\begin{figure}[H]
    \centering
    \includegraphics[width=0.75\linewidth]{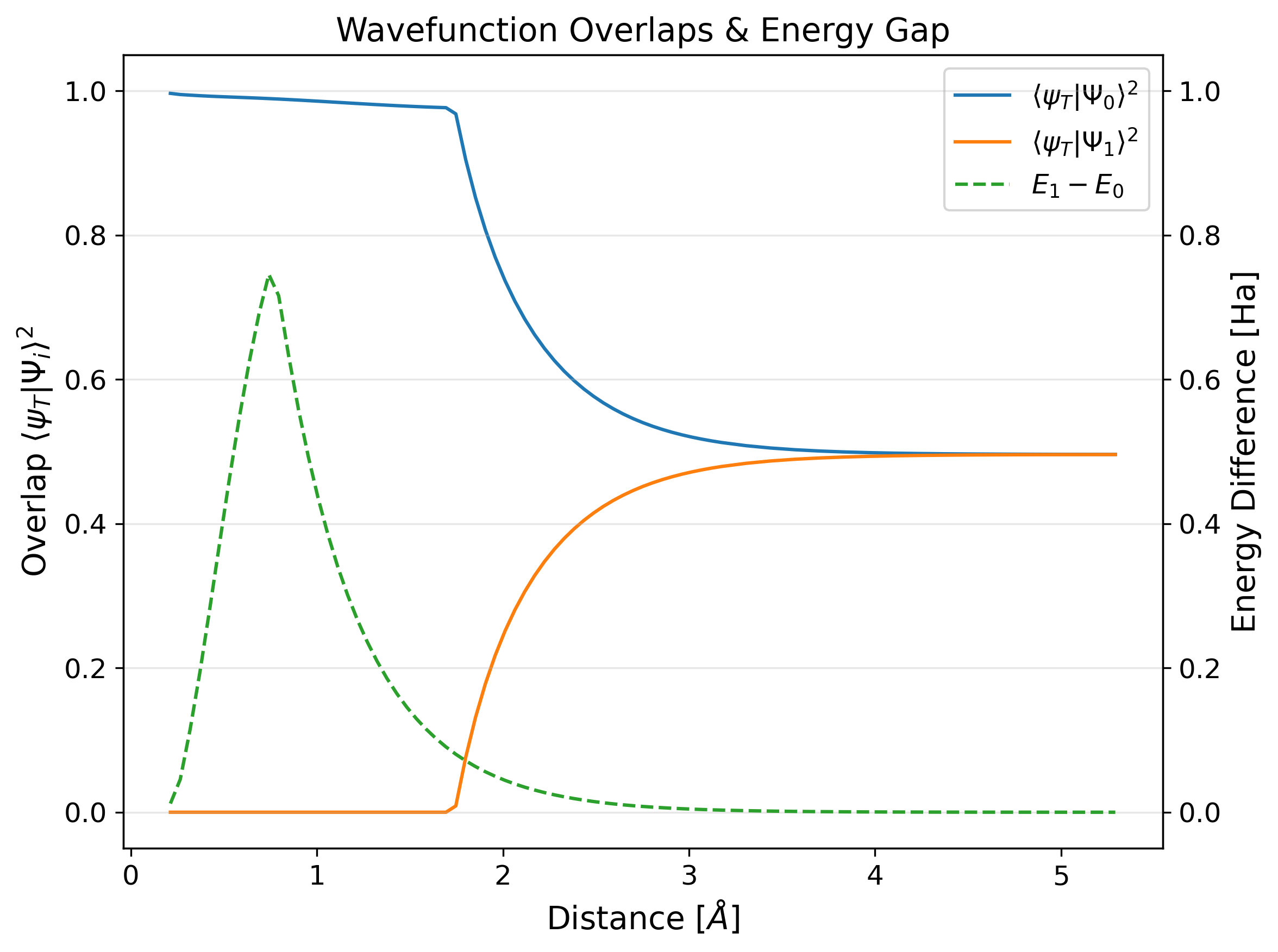}
    \caption{\footnotesize For the dissociation of He$_2^+$, the energy gap between the lowest and first-excited doublet eigenstates (right vertical axis) and the squared overlap of the lowest-energy UHF trial wavefunction with the two eigenstates (left vertical axis) are shown.   
    }
    \label{fig:He2pOverlapFCI}
\end{figure}

In the spirit of understanding the unexpected imaginary-time trajectory that results from use of the UHF II state as the trial wavefunction, 
Figure \ref{fig:He2pOverlapFCI} shows the energy gap between the two lowest-energy exact eigenstates, $\Psi_0$ and $\Psi_1$, and their overlaps with the stable UHF II trial wavefunction, as a function of bond distance.  
At very short distance, $\Psi_0$ and $\Psi_1$ are nearly degenerate but then separate to a maximum of roughly 0.754 Ha at 0.76 \AA.  As the bond distance continues to increase, the energy gap decreases rapidly to again approach zero.  Before 2 \AA, the squared overlap of the UHF II trial with the exact ground-state, $|\langle \psi_{\text{UHF II}}|\Psi_0\rangle|^2$, is close to one; just before 2\AA \ this quantity starts to drop while $|\langle\psi_{\text{UHF II}}|\Psi_1\rangle|^2$ increases correspondingly; by 4\AA \ both squared overlaps are roughly equal at 0.5.  The abnormal ph-AFQMC trajectories, which are observed at roughly 2\AA \ and beyond, would seem to coincide with the onset of a non-zero overlap of the trial with the first excited doublet eigenstate, and when $\Psi_0$ and $\Psi_1$ become nearly-degenerate in energy.  Indeed, as the He atoms separate, the two eigenstates are $\pm$ combinations of the two UHF states with the hole localized on different He atoms; it is reasonable to expect that an importance function that consists of only a single configuration and does not encode knowledge of the correct relative phase in the ground eigenstate expansion could lead to suboptimal sampling in ph-AFQMC.

The overly negative ph-AFQMC energy in the intermediate region with the UHF II trial must be due to overly negative walker local energies.  Consider a walker which starts out close to the UHF II trial wavefunction, with the hole localized on He atom A -- a state we will denote He$_\text{A}$.  As the AFQMC propagation is simply a string of one-body operators, it will be very probable for one or more walker to evolve to have the hole localized on atom B (He$_\text{B}$).  The walker's local energy in this situation becomes:
\begin{equation}
        E_{\text{L}} = \frac{\langle \text{He}_\text{A} | \hat{H} | \text{He}_\text{B} \rangle}{\langle \text{He}_\text{A} | \text{He}_\text{B} \rangle}.
\end{equation}
At intermediate distances, the denominator will be small and the numerator will be negative, which would account for the overly negative local energies observed.  Furthermore, since the weights are proportional to $e^{-\Delta\tau E_{\text{L}}}$, the population control algorithm -- which periodically duplicates walkers with large weights and removes walkers with small weights -- will accelerate the growth of a large population of walkers with determinants in the He$_B$ state.  
Indeed, this behavior is confirmed in panel (b) of Figure \ref{fig:he2_afqmc_energy_overlaps}, where the overlap of the ph-AFQMC wave function with the UHF II trial at $R= 4.62 \text{\AA}$ collapses while the overlap  with $\text{UHF}_B$ rises, coinciding with when the energy trajectory drops toward the second plateau.
In contrast, for $R= 1.50 \text{\AA}$, the ph-AFQMC wave function 
retains a large overlap with the UHF II trial and the energy 
trajectory is non-pathological.

\begin{figure}[H]
    \centering
    \includegraphics[width=0.75\linewidth]{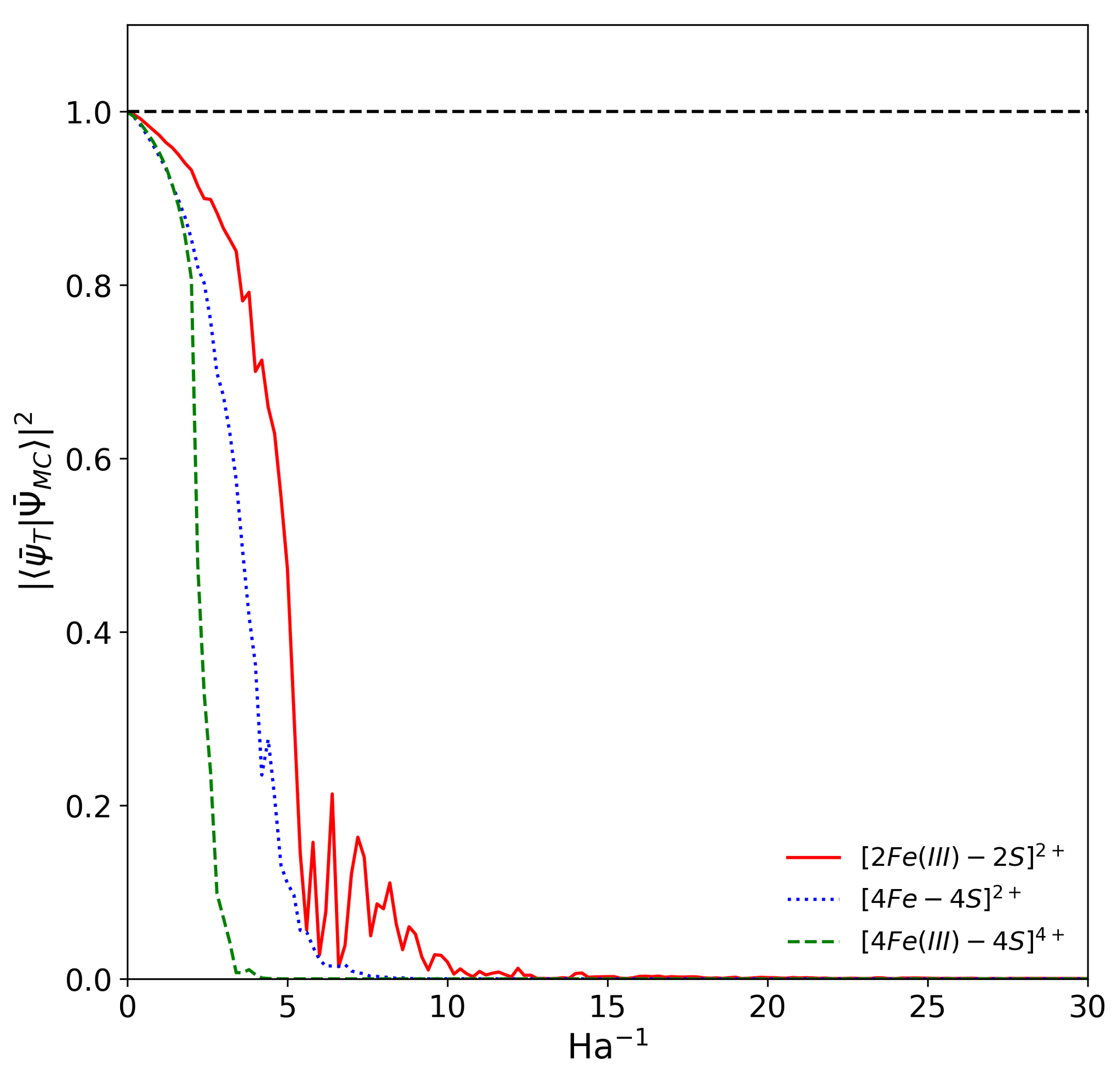}
    \caption{\footnotesize Normalized squared overlaps between optimal-SDM UHF trial and the respective ph-AFQMC wave function for each iron-sulfur cluster studied in this work.}
    \label{fig:fe_s_overlaps}
\end{figure}

To verify whether or not similar walker dynamics are at play for the three iron-sulfur clusters, we computed the 
overlaps of the ph-AFQMC wave functions with the SDM-optimal UHF trials discussed above. As shown in Figure \ref{fig:fe_s_overlaps}, the overlaps decay to zero in all cases very early in the imaginary time propagation, well before the energy equilibrates.
While these UHF trials led to very accurate ph-AFQMC energies, they evidently do not effectively constrain the walkers to retain a high overlap with the trial, unlike the well-behaved example of He$_2^+$ at $1.50 \text{\AA}$. 
The multitude of possible UHF states for the iron-sulfur clusters makes determining which of them, if any, might have a growing 
or large overlap with the AFQMC wave function challenging and will be the focus of future investigation.

Nevertheless, the decay of the overlap between the stochastically evolving wavefunction and the UHF trial suggests that, on average, small local energy denominators are causing overly negative local energies.  This is a sampling bias that is distinct from the phaseless bias (indeed, they likely bias the energy in opposite directions).  
The dominance of this sampling bias would explain the overly negative ph-AFQMC energies found in our study of [4Fe--4S]$^{4+}$, and also previous literature reports of ph-AFQMC overcorrelating at stretched bond lengths (e.g., in N$_2$, C$_2$, and H$_{10}$) when UHF trials are used,\cite{al2007bond,lee_twenty_2022,motta2017towards} and the convergence ``from below'' in the [2Fe--2S] model from Ref. \citenum{huang2024gpu} as the number of determinants in the trial was increased. 
The overlap results imply that this sampling bias is at play even in [2Fe--2S]$^{2+}$ and [4Fe--4S]$^{2+}$ where the optimal-SDM UHF trials led to ph-AFQMC energies above or exactly on top of the theoretical best estimates, albeit possibly to a lesser extent given that the energy trajectories equilibrate monotonically and are relatively well-behaved.  As a final point, our present findings are consistent with the results of Kjønstad et al.\cite{kjonstad2025systematic}, who have found that systematically improving UHF trials with, e.g., CI subspace-projected UCCSD makes the total ph-AFQMC energies more positive for various Fe-S systems.

\section{Conclusions}

In this work we have considered three Fe-S active space models derived from synthetic, biomimetic clusters -- [2Fe--2S]$^{2+}$, [4Fe--4S]$^{2+}$, and [4Fe--4S]$^{4+}$ -- that capture the valence (Fe 3$d$ and S 3$p$) electron correlation effects at play in many biological metalloenzymes.    
These systems represent a frontier in the field of electronic structure theory, as the delicate entanglement present in the ground and other low-lying electronic states is challenging to reliably describe with conventional methods such as approximate density functionals and single-reference perturbation theories.  We show that in each case there exists a UHF solution that can lead to reasonably accurate ph-AFQMC and CCSD(T) ground-state energies vs.\@ the theoretical best estimates, and highlight the importance of the choice of trial or reference UHF state.  Importantly, the lowest-energy UHF state does not always lead to the most accurate ph-AFQMC and CCSD(T) energies. When informed by physical symmetries and ideally structural or spectroscopic characterization of the metal oxidation states and relative spin orientations, we demonstrate that a function of the atomic Mulliken spin densities -- which we call the spin density metric -- can reliably identify the optimal UHF trial/reference state for subsequent ph-AFQMC/CCSD(T) calculations.  This choice enables very high accuracy in specific cases:  for one spin isomer in [4Fe--4S]$^{2+}$ ph-AFQMC is essentially exact, and CCSD(T) is off by a mere 3.6 mHa for [4Fe--4S]$^{4+}$.  However, across all of the Fe-S systems investigated in this work the performance of ph-AFQMC with optimal-SDM UHF trials is more modest, showing average errors of $\sim$10-20 mHa from the theoretical best estimates.  Nevertheless given the low $\mathcal{O}$($N^4$) scaling of ph-AFQMC-UHF (per-sample) with system size, vs the $\mathcal{O}$($N^7$) and $\mathcal{O}$($e^N$) scaling of CCSD(T) and methods such as selected CI and DMRG, respectively, this level of accuracy for such challenging strongly correlated systems is encouraging.  Future work is needed to explore different geometries\cite{mejuto2022effect} and charge states of these and other Fe-S clusters\cite{skeel2025iron}, in addition to other types of polynuclear transition metal systems.

The unexpected utility of UHF trials in ph-AFQMC, even for strongly correlated electronic states, has a number of exciting implications.   
UHF has many desirable formal properties, e.g., it is orbital invariant (to occupied-occupied and virtual-virtual rotations), entails a relatively inexpensive $\mathcal{O}(N^4)$ computational cost scaling with system size, and is size-extensive.  The latter is highly desirable (in fact, necessary) to produce meaningful results for extended systems in the thermodynamic limit.  Furthermore, in our experience, the fact that an active space selection is not required and the relatively small space of optimization parameters involved in the UHF SCF procedure relative to methods such as CASSCF, are welcome simplifications. From a chemical perspective, the effective one-electron picture of mean-field theory permits natural interpretations in terms of beloved (and useful) chemical concepts such as orbital shapes/energies, bond orders, and oxidation states, grounded by a constantly advancing array of experimental techniques (EPR, Mossbauer, X-ray photoelectron, NMR, etc.).

However, a closer look at the ph-AFQMC wavefunctions reveals an unexpected deficiency in the ability of even the optimal UHF state to act as an effective importance function in the case of these three Fe-S systems. Specifically, the average overlap of the stochastic QMC wavefunction with the UHF trial was found to decay rapidly to zero, as was found also for the stretched helium dimer cation.  The small local energy denominators result in pathological reweighting that leads eventually to an equilibrated walker population with zero overlap with the UHF trial wavefunction, which tends to push the ph-AFQMC energy estimate too negative.  In the cases of stretched He$_2^+$ and [4Fe--4S]$^{4+}$ we find non-monotonic energy trajectories in imaginary time resulting in an overly negative ph-AFQMC energy, whereas in [2Fe--2S]$^{2+}$ and [4Fe--4S]$^{4+}$ the trajectories are not unusual and the artificial overcorrelation due to this sampling bias appears to be compensated (sometimes exactly) by the phaseless bias.  

These observations have fundamental implications for how we can understand the role of the trial wavefunction in ph-AFQMC, and moreover how we might design improved classes of trial wavefunctions (at present using polynomial-cost classical algorithms but in the future potentially exploiting quantum computation/simulation\cite{huggins_unbiasing_2022,amsler2023classical,huang2024evaluating,zhao2025quantum,danilov_enhancing_2025,yoshida2025auxiliary}).  One takeaway is that when the true ground-state involves multiple large-weighted configurations superposed with specific relative phases (which, as most clearly seen in the case of stretched He$_2^+$, can distinguish one low-energy eigenstate from another), the determinants, their phase relationships, and moreover the requisite physical symmetries ought to be encoded in the trial wavefunction for properly-functioning importance sampling to be realized.  There is also the (admittedly riskier, but nevertheless compelling) possibility that highly accurate energies can be achieved with suitable mean-field wavefunctions despite a significant sampling bias as long as there is adequate cancellation of error, e.g., from the phaseless bias -- though much more work is required to explore and test the robustness of such a strategy.  In our view, the choice of trial wavefunction of any type, whether correlated or not, ought to consider not just energies but also properties that can be deduced from physical symmetries (e.g. group theoretic arguments) and/or experimental measurements. It is easy to imagine alternatives to the spin density metric proposed here for UHF states that involve, e.g., spin-spin correlation functions, local $\hat{S}_z$ expectation values, etc.  It is an open question how to choose and systematically improve an optimal trial wavefunction \emph{a priori} for strongly correlated states (especially of large system sizes), though ideas involving, e.g., self-consistency\cite{qin2016coupling,sukurma2025self} and the breaking and/or restoring of a hierarchy of symmetries\cite{shi2014symmetry,ung2026study} appear promising.

\section{Acknowledgements}
We acknowledge helpful discussions with Garnet Chan, Eirik Kjønstad, Mario Motta, Sandeep Sharma, and Shiwei Zhang, and thank the Chan group for providing us with the integrals for the [4Fe(III)--4S]$^{4+}$ cluster.
This work was supported in part by the Big-Data Private-Cloud Research Cyberinfrastructure MRI-award funded by NSF under grant CNS-1338099 and by Rice University's Center for Research Computing (CRC). This research also used resources of the Oak Ridge Leadership Computing Facility, which is a DOE Office of Science User Facility supported under Contract DE-AC05-00OR22725. B. Ganoe is a Welch Postdoctoral Fellow of The Life Sciences
Research Foundation, Award No. R-C-0002-20240404.
J. Shee acknowledges support from the Robert A. Welch
Foundation, Award Number C-2212.

\section{Supplementary information}
UHF trial energies, $\langle \hat{S}^2\rangle$, spin density analysis; ph-AFQMC, CCSD, and CCSD(T) energies

\bibliography{main,don_references}

@article{mishra_study_2022,
	title = {Study of {Self}-{Interaction} {Errors} in {Density} {Functional} {Calculations} of {Magnetic} {Exchange} {Coupling} {Constants} {Using} {Three} {Self}-{Interaction} {Correction} {Methods}},
	volume = {126},
	copyright = {https://doi.org/10.15223/policy-029},
	issn = {1089-5639, 1520-5215},
	url = {https://pubs.acs.org/doi/10.1021/acs.jpca.1c10354},
	doi = {10.1021/acs.jpca.1c10354},
	language = {en},
	number = {12},
	urldate = {2026-04-24},
	journal = {The Journal of Physical Chemistry A},
	author = {Mishra, Prakash and Yamamoto, Yoh and Chang, Po-Hao and Nguyen, Duyen B. and Peralta, Juan E. and Baruah, Tunna and Zope, Rajendra R.},
	month = mar,
	year = {2022},
	pages = {1923--1935},
}

@misc{eirik_f_kjonstad_can_2026,
	title = {Can phaseless auxiliary-field quantum {Monte} {Carlo} describe iron-sulfur clusters?},
	publisher = {arXiv},
	author = {{Eirik F. Kjønstad} and {Huanchen Zhai} and {James Shee} and {Sandeep Sharma} and {Garnet Kin-Lic Chan}},
	year = {2026},
}

@misc{song_spin-adapted_2026,
	title = {Spin-{Adapted} {Restricted} {Open}-{Shell} {Hartree}-{Fock} and {Its} {Dynamic} {Correlation} {Extension}},
	url = {https://chemrxiv.org/doi/full/10.26434/chemrxiv.15000528/v2},
	doi = {10.26434/chemrxiv.15000528/v2},
	language = {en},
	urldate = {2026-03-06},
	author = {Song, Maru and Bonfirraro, Luca and Fdez. Galván, Ignacio and Lindh, Roland and Li Manni, Giovanni},
	month = mar,
	year = {2026},
}

@article{li_electronic_2019,
	title = {The electronic complexity of the ground-state of the {FeMo} cofactor of nitrogenase as relevant to quantum simulations},
	volume = {150},
	issn = {0021-9606, 1089-7690},
	url = {https://pubs.aip.org/jcp/article/150/2/024302/197301/The-electronic-complexity-of-the-ground-state-of},
	doi = {10.1063/1.5063376},
	language = {en},
	number = {2},
	urldate = {2025-10-01},
	journal = {The Journal of Chemical Physics},
	author = {Li, Zhendong and Li, Junhao and Dattani, Nikesh S. and Umrigar, C. J. and Chan, Garnet Kin-Lic},
	month = jan,
	year = {2019},
	pages = {024302},
}

@article{reiher_elucidating_2017,
	title = {Elucidating reaction mechanisms on quantum computers},
	volume = {114},
	issn = {0027-8424, 1091-6490},
	url = {https://pnas.org/doi/full/10.1073/pnas.1619152114},
	doi = {10.1073/pnas.1619152114},
	language = {en},
	number = {29},
	urldate = {2026-02-25},
	journal = {Proceedings of the National Academy of Sciences},
	author = {Reiher, Markus and Wiebe, Nathan and Svore, Krysta M. and Wecker, Dave and Troyer, Matthias},
	month = jul,
	year = {2017},
	pages = {7555--7560},
}

@article{alexeev_perspective_2025,
	title = {A {Perspective} on {Quantum} {Computing} {Applications} in {Quantum} {Chemistry} {Using} 25–100 {Logical} {Qubits}},
	volume = {21},
	copyright = {https://doi.org/10.15223/policy-029},
	issn = {1549-9618, 1549-9626},
	url = {https://pubs.acs.org/doi/10.1021/acs.jctc.5c01038},
	doi = {10.1021/acs.jctc.5c01038},
	language = {en},
	number = {22},
	urldate = {2026-02-25},
	journal = {Journal of Chemical Theory and Computation},
	author = {Alexeev, Yuri and Batista, Victor S. and Bauman, Nicholas and Bertels, Luke and Claudino, Daniel and Dutta, Rishab and Gagliardi, Laura and Godwin, Scott and Govind, Niranjan and Head-Gordon, Martin and Hermes, Matthew R. and Kowalski, Karol and Li, Ang and Liu, Chenxu and Liu, Junyu and Liu, Ping and García-Lastra, Juan M. and Mejia-Rodriguez, Daniel and Mueller, Karl and Otten, Matthew and Peng, Bo and Raugas, Mark and Reiher, Markus and Rigor, Paul and Shaw, Wendy J. and Van Schilfgaarde, Mark and Vegge, Tejs and Zhang, Yu and Zheng, Muqing and Zhu, Linghua},
	month = nov,
	year = {2025},
	pages = {11335--11357},
}

@article{grunwald_vibrational_2025,
	title = {Vibrational {Architecture} of [{Fe}$_{\textrm{4}}$ {S}$_{\textrm{4}}$ ]$^{\textrm{0/1+/2+/3+/4+}}$ {Iron}–{Sulfur} {Cubanes}},
	volume = {64},
	copyright = {https://doi.org/10.15223/policy-029},
	issn = {0020-1669, 1520-510X},
	url = {https://pubs.acs.org/doi/10.1021/acs.inorgchem.5c02572},
	doi = {10.1021/acs.inorgchem.5c02572},
	language = {en},
	number = {36},
	urldate = {2026-02-24},
	journal = {Inorganic Chemistry},
	author = {Grunwald, Liam and Pelmenschikov, Vladimir and Wang, Hongxin and Yoda, Yoshitaka and Nagasawa, Nobumoto and Tamasaku, Kenji and Cramer, Stephen P. and Mougel, Victor},
	month = sep,
	year = {2025},
	pages = {18062--18067},
}

@article{ohki_synthetic_2011,
	title = {Synthetic analogues of [{Fe}$_{\textrm{4}}$ {S}$_{\textrm{4}}$ ({Cys})$_{\textrm{3}}$ ({His})] in hydrogenases and [{Fe}$_{\textrm{4}}$ {S}$_{\textrm{4}}$ ({Cys})$_{\textrm{4}}$ ] in {HiPIP} derived from all-ferric [{Fe}$_{\textrm{4}}$ {S}$_{\textrm{4}}$ \{{N}({SiMe}$_{\textrm{3}}$ )$_{\textrm{2}}$ \}$_{\textrm{4}}$ ]},
	volume = {108},
	issn = {0027-8424, 1091-6490},
	url = {https://pnas.org/doi/full/10.1073/pnas.1106472108},
	doi = {10.1073/pnas.1106472108},
	language = {en},
	number = {31},
	urldate = {2026-02-24},
	journal = {Proceedings of the National Academy of Sciences},
	author = {Ohki, Yasuhiro and Tanifuji, Kazuki and Yamada, Norihiro and Imada, Motosuke and Tajima, Tomoyuki and Tatsumi, Kazuyuki},
	month = aug,
	year = {2011},
	pages = {12635--12640},
}

@article{robledo-moreno_chemistry_2025,
	title = {Chemistry beyond the scale of exact diagonalization on a quantum-centric supercomputer},
	volume = {11},
	issn = {2375-2548},
	url = {https://www.science.org/doi/10.1126/sciadv.adu9991},
	doi = {10.1126/sciadv.adu9991},
	language = {en},
	number = {25},
	urldate = {2026-02-18},
	journal = {Science Advances},
	author = {Robledo-Moreno, Javier and Motta, Mario and Haas, Holger and Javadi-Abhari, Ali and Jurcevic, Petar and Kirby, William and Martiel, Simon and Sharma, Kunal and Sharma, Sandeep and Shirakawa, Tomonori and Sitdikov, Iskandar and Sun, Rong-Yang and Sung, Kevin J. and Takita, Maika and Tran, Minh C. and Yunoki, Seiji and Mezzacapo, Antonio},
	month = jun,
	year = {2025},
	pages = {eadu9991},
}

@incollection{bezkorovainy_iron-sulfur_1980,
	address = {Boston, MA},
	title = {The {Iron}-{Sulfur} {Proteins}},
	isbn = {978-1-4684-3781-2 978-1-4684-3779-9},
	url = {http://link.springer.com/10.1007/978-1-4684-3779-9_8},
	doi = {10.1007/978-1-4684-3779-9_8},
	language = {en},
	urldate = {2025-12-22},
	booktitle = {Biochemistry of {Nonheme} {Iron}},
	publisher = {Springer US},
	author = {Bezkorovainy, Anatoly},
	collaborator = {Bezkorovainy, Anatoly},
	year = {1980},
	pages = {343--393},
}

@article{shee_achieving_2019,
	title = {On {Achieving} {High} {Accuracy} in {Quantum} {Chemical} {Calculations} of 3d {Transition} {Metal}-{Containing} {Systems}: {A} {Comparison} of {Auxiliary}-{Field} {Quantum} {Monte} {Carlo} with {Coupled} {Cluster}, {Density} {Functional} {Theory}, and {Experiment} for {Diatomic} {Molecules}},
	volume = {15},
	issn = {1549-9618, 1549-9626},
	shorttitle = {On {Achieving} {High} {Accuracy} in {Quantum} {Chemical} {Calculations} of 3 \textit{d} {Transition} {Metal}-{Containing} {Systems}},
	url = {https://pubs.acs.org/doi/10.1021/acs.jctc.9b00083},
	doi = {10.1021/acs.jctc.9b00083},
	language = {en},
	number = {4},
	urldate = {2023-09-11},
	journal = {Journal of Chemical Theory and Computation},
	author = {Shee, James and Rudshteyn, Benjamin and Arthur, Evan J. and Zhang, Shiwei and Reichman, David R. and Friesner, Richard A.},
	month = apr,
	year = {2019},
	pages = {2346--2358},
}

@article{johnson_structure_2005,
	title = {Structure, {Function}, and {Formation} of biological iron-sulfur clusters},
	volume = {74},
	issn = {0066-4154, 1545-4509},
	url = {https://www.annualreviews.org/doi/10.1146/annurev.biochem.74.082803.133518},
	doi = {10.1146/annurev.biochem.74.082803.133518},
	language = {en},
	number = {1},
	urldate = {2025-07-18},
	journal = {Annual Review of Biochemistry},
	publisher = {Annual Reviews},
	author = {Johnson, Deborah C. and Dean, Dennis R. and Smith, Archer D. and Johnson, Michael K.},
	month = jun,
	year = {2005},
	pages = {247--281},
}

@article{danilov_enhancing_2025,
	title = {Enhancing the {Accuracy} and {Efficiency} of {Sample}-{Based} {Quantum} {Diagonalization} with {Phaseless} {Auxiliary}-{Field} {Quantum} {Monte} {Carlo}},
	volume = {21},
	copyright = {https://doi.org/10.15223/policy-029},
	issn = {1549-9618, 1549-9626},
	url = {https://pubs.acs.org/doi/10.1021/acs.jctc.5c01407},
	doi = {10.1021/acs.jctc.5c01407},
	language = {en},
	number = {22},
	urldate = {2025-12-11},
	journal = {Journal of Chemical Theory and Computation},
	author = {Danilov, Don and Robledo-Moreno, Javier and Sung, Kevin J. and Motta, Mario and Shee, James},
	month = nov,
	year = {2025},
	pages = {11585--11594},
}

@article{mahajan_selected_2022,
	title = {Selected configuration interaction wave functions in phaseless auxiliary field quantum {Monte} {Carlo}},
	volume = {156},
	issn = {0021-9606, 1089-7690},
	url = {https://pubs.aip.org/jcp/article/156/17/174111/2841175/Selected-configuration-interaction-wave-functions},
	doi = {10.1063/5.0087047},
	language = {en},
	number = {17},
	urldate = {2024-07-29},
	journal = {The Journal of Chemical Physics},
	author = {Mahajan, Ankit and Lee, Joonho and Sharma, Sandeep},
	month = may,
	year = {2022},
	pages = {174111},
}

@article{tanifuji_metalsulfur_2020,
	title = {Metal–{Sulfur} {Compounds} in {N}$_{\textrm{2}}$ {Reduction} and {Nitrogenase}-{Related} {Chemistry}},
	volume = {120},
	copyright = {https://doi.org/10.15223/policy-029},
	issn = {0009-2665, 1520-6890},
	url = {https://pubs.acs.org/doi/10.1021/acs.chemrev.9b00544},
	doi = {10.1021/acs.chemrev.9b00544},
	language = {en},
	number = {12},
	urldate = {2025-07-18},
	journal = {Chemical Reviews},
	publisher = {American Chemical Society (ACS)},
	author = {Tanifuji, Kazuki and Ohki, Yasuhiro},
	month = jun,
	year = {2020},
	pages = {5194--5251},
}

@article{beinert_iron-sulfur_1997,
	title = {Iron-{Sulfur} {Clusters}: {Nature}'s {Modular}, {Multipurpose} {Structures}},
	volume = {277},
	issn = {0036-8075, 1095-9203},
	shorttitle = {Iron-{Sulfur} {Clusters}},
	url = {https://www.science.org/doi/10.1126/science.277.5326.653},
	doi = {10.1126/science.277.5326.653},
	language = {en},
	number = {5326},
	urldate = {2025-07-18},
	journal = {Science},
	publisher = {American Association for the Advancement of Science (AAAS)},
	author = {Beinert, Helmut and Holm, Richard H. and Münck, Eckard},
	month = aug,
	year = {1997},
	pages = {653--659},
}

@article{gupta_ironsulfur_2020,
	title = {Iron–sulfur cluster signaling: {The} common thread in fungal iron regulation},
	volume = {55},
	copyright = {https://www.elsevier.com/tdm/userlicense/1.0/},
	issn = {1367-5931},
	shorttitle = {Iron–sulfur cluster signaling},
	url = {https://linkinghub.elsevier.com/retrieve/pii/S1367593120300247},
	doi = {10.1016/j.cbpa.2020.02.008},
	language = {en},
	urldate = {2025-07-18},
	journal = {Current Opinion in Chemical Biology},
	publisher = {Elsevier BV},
	author = {Gupta, Malini and Outten, Caryn E.},
	month = apr,
	year = {2020},
	pages = {189--201},
}

@article{read_mitochondrial_2021,
	title = {Mitochondrial iron–sulfur clusters: {Structure}, function, and an emerging role in vascular biology},
	volume = {47},
	copyright = {https://www.elsevier.com/tdm/userlicense/1.0/},
	issn = {2213-2317},
	shorttitle = {Mitochondrial iron–sulfur clusters},
	url = {https://linkinghub.elsevier.com/retrieve/pii/S2213231721003244},
	doi = {10.1016/j.redox.2021.102164},
	language = {en},
	urldate = {2025-07-18},
	journal = {Redox Biology},
	publisher = {Elsevier BV},
	author = {Read, Austin D. and Bentley, Rachel Et. and Archer, Stephen L. and Dunham-Snary, Kimberly J.},
	month = nov,
	year = {2021},
	pages = {102164},
}

@article{sharma_low-energy_2014,
	title = {Low-energy spectrum of iron–sulfur clusters directly from many-particle quantum mechanics},
	volume = {6},
	issn = {1755-4330, 1755-4349},
	url = {https://www.nature.com/articles/nchem.2041},
	doi = {10.1038/nchem.2041},
	language = {en},
	number = {10},
	urldate = {2024-11-12},
	journal = {Nature Chemistry},
	author = {Sharma, Sandeep and Sivalingam, Kantharuban and Neese, Frank and Chan, Garnet Kin-Lic},
	month = oct,
	year = {2014},
	pages = {927--933},
}

@article{marti-dafcik_spin_2025,
	title = {Spin coupling is all you need: {Encoding} strong electron correlation in molecules on quantum computers},
	volume = {7},
	copyright = {https://creativecommons.org/licenses/by/4.0/},
	issn = {2643-1564},
	shorttitle = {Spin coupling is all you need},
	url = {https://link.aps.org/doi/10.1103/PhysRevResearch.7.013191},
	doi = {10.1103/physrevresearch.7.013191},
	language = {en},
	number = {1},
	urldate = {2025-07-14},
	journal = {Physical Review Research},
	publisher = {American Physical Society (APS)},
	author = {Marti-Dafcik, Daniel and Burton, Hugh G. A. and Tew, David P.},
	month = feb,
	year = {2025},
}

@article{lee_evaluating_2023,
	title = {Evaluating the evidence for exponential quantum advantage in ground-state quantum chemistry},
	volume = {14},
	copyright = {https://creativecommons.org/licenses/by/4.0},
	issn = {2041-1723},
	url = {https://www.nature.com/articles/s41467-023-37587-6},
	doi = {10.1038/s41467-023-37587-6},
	language = {en},
	number = {1},
	urldate = {2025-07-10},
	journal = {Nature Communications},
	publisher = {Springer Science and Business Media LLC},
	author = {Lee, Seunghoon and Lee, Joonho and Zhai, Huanchen and Tong, Yu and Dalzell, Alexander M. and Kumar, Ashutosh and Helms, Phillip and Gray, Johnnie and Cui, Zhi-Hao and Liu, Wenyuan and Kastoryano, Michael and Babbush, Ryan and Preskill, John and Reichman, David R. and Campbell, Earl T. and Valeev, Edward F. and Lin, Lin and Chan, Garnet Kin-Lic},
	month = apr,
	year = {2023},
}

@article{burton_energy_2021,
	title = {Energy {Landscapes} for {Electronic} {Structure}},
	volume = {17},
	copyright = {https://doi.org/10.15223/policy-029},
	issn = {1549-9618, 1549-9626},
	url = {https://pubs.acs.org/doi/10.1021/acs.jctc.0c00772},
	doi = {10.1021/acs.jctc.0c00772},
	language = {en},
	number = {1},
	urldate = {2025-03-27},
	journal = {Journal of Chemical Theory and Computation},
	author = {Burton, Hugh G. A. and Wales, David J.},
	month = jan,
	year = {2021},
	pages = {151--169},
}

@article{jiang_unbiasing_2025,
	title = {Unbiasing fermionic auxiliary-field quantum {Monte} {Carlo} with matrix product state trial wavefunctions},
	volume = {7},
	issn = {2643-1564},
	url = {https://link.aps.org/doi/10.1103/PhysRevResearch.7.013038},
	doi = {10.1103/PhysRevResearch.7.013038},
	language = {en},
	number = {1},
	urldate = {2025-03-13},
	journal = {Physical Review Research},
	author = {Jiang, Tong and O'Gorman, Bryan and Mahajan, Ankit and Lee, Joonho},
	month = jan,
	year = {2025},
	pages = {013038},
}

@article{mahajan_beyond_2025,
	title = {Beyond {CCSD}({T}) {Accuracy} at {Lower} {Scaling} with {Auxiliary} {Field} {Quantum} {Monte} {Carlo}},
	volume = {21},
	copyright = {https://doi.org/10.15223/policy-029},
	issn = {1549-9618, 1549-9626},
	url = {https://pubs.acs.org/doi/10.1021/acs.jctc.4c01314},
	doi = {10.1021/acs.jctc.4c01314},
	language = {en},
	number = {4},
	urldate = {2025-03-13},
	journal = {Journal of Chemical Theory and Computation},
	author = {Mahajan, Ankit and Thorpe, James H. and Kurian, Jo S. and Reichman, David R. and Matthews, Devin A. and Sharma, Sandeep},
	month = feb,
	year = {2025},
	pages = {1626--1642},
}

@article{li_spin-projected_2017,
	title = {Spin-{Projected} {Matrix} {Product} {States}: {Versatile} {Tool} for {Strongly} {Correlated} {Systems}},
	volume = {13},
	issn = {1549-9618, 1549-9626},
	shorttitle = {Spin-{Projected} {Matrix} {Product} {States}},
	url = {https://pubs.acs.org/doi/10.1021/acs.jctc.7b00270},
	doi = {10.1021/acs.jctc.7b00270},
	language = {en},
	number = {6},
	urldate = {2025-03-04},
	journal = {Journal of Chemical Theory and Computation},
	author = {Li, Zhendong and Chan, Garnet Kin-Lic},
	month = jun,
	year = {2017},
	pages = {2681--2695},
}

@article{ganoe_notion_2024,
	title = {On the notion of strong correlation in electronic structure theory},
	issn = {1359-6640, 1364-5498},
	url = {http://pubs.rsc.org/en/Content/ArticleLanding/2024/FD/D4FD00066H},
	doi = {10.1039/D4FD00066H},
	language = {en},
	urldate = {2024-04-18},
	journal = {Faraday Discussions},
	author = {Ganoe, Brad and Shee, James},
	year = {2024},
	pages = {10.1039.D4FD00066H},
}

@article{malone_ipie_2023,
	title = {ipie : {A} {Python}-{Based} {Auxiliary}-{Field} {Quantum} {Monte} {Carlo} {Program} with {Flexibility} and {Efficiency} on {CPUs} and {GPUs}},
	volume = {19},
	issn = {1549-9618, 1549-9626},
	shorttitle = {ipie},
	url = {https://pubs.acs.org/doi/10.1021/acs.jctc.2c00934},
	doi = {10.1021/acs.jctc.2c00934},
	language = {en},
	number = {1},
	urldate = {2023-09-04},
	journal = {Journal of Chemical Theory and Computation},
	author = {Malone, Fionn D. and Mahajan, Ankit and Spencer, James S. and Lee, Joonho},
	month = jan,
	year = {2023},
	pages = {109--121},
}

@article{shi_symmetry_2013,
	title = {Symmetry in auxiliary-field quantum {Monte} {Carlo} calculations},
	volume = {88},
	issn = {1098-0121, 1550-235X},
	url = {https://link.aps.org/doi/10.1103/PhysRevB.88.125132},
	doi = {10.1103/PhysRevB.88.125132},
	language = {en},
	number = {12},
	urldate = {2024-01-31},
	journal = {Physical Review B},
	author = {Shi, Hao and Zhang, Shiwei},
	month = sep,
	year = {2013},
	pages = {125132},
}

@article{motta_ab_2018,
	title = {Ab initio computations of molecular systems by the auxiliary‐field quantum {Monte} {Carlo} method},
	volume = {8},
	issn = {1759-0876, 1759-0884},
	url = {https://wires.onlinelibrary.wiley.com/doi/10.1002/wcms.1364},
	doi = {10.1002/wcms.1364},
	language = {en},
	number = {5},
	urldate = {2023-09-11},
	journal = {WIREs Computational Molecular Science},
	author = {Motta, Mario and Zhang, Shiwei},
	month = sep,
	year = {2018},
	pages = {e1364},
}

@article{izsak_measuring_2023,
	title = {Measuring {Electron} {Correlation}: {The} {Impact} of {Symmetry} and {Orbital} {Transformations}},
	volume = {19},
	issn = {1549-9618, 1549-9626},
	shorttitle = {Measuring {Electron} {Correlation}},
	url = {https://pubs.acs.org/doi/10.1021/acs.jctc.3c00122},
	doi = {10.1021/acs.jctc.3c00122},
	language = {en},
	number = {10},
	urldate = {2023-10-09},
	journal = {Journal of Chemical Theory and Computation},
	author = {Izsák, Róbert and Ivanov, Aleksei V. and Blunt, Nick S. and Holzmann, Nicole and Neese, Frank},
	month = may,
	year = {2023},
	pages = {2703--2720},
}

@article{huggins_unbiasing_2022,
	title = {Unbiasing fermionic quantum {Monte} {Carlo} with a quantum computer},
	volume = {603},
	issn = {0028-0836, 1476-4687},
	url = {https://www.nature.com/articles/s41586-021-04351-z},
	doi = {10.1038/s41586-021-04351-z},
	language = {en},
	number = {7901},
	urldate = {2023-09-12},
	journal = {Nature},
	author = {Huggins, William J. and O’Gorman, Bryan A. and Rubin, Nicholas C. and Reichman, David R. and Babbush, Ryan and Lee, Joonho},
	month = mar,
	year = {2022},
	pages = {416--420},
}

@article{lee_twenty_2022,
	title = {Twenty {Years} of {Auxiliary}-{Field} {Quantum} {Monte} {Carlo} in {Quantum} {Chemistry}: {An} {Overview} and {Assessment} on {Main} {Group} {Chemistry} and {Bond}-{Breaking}},
	volume = {18},
	issn = {1549-9618, 1549-9626},
	shorttitle = {Twenty {Years} of {Auxiliary}-{Field} {Quantum} {Monte} {Carlo} in {Quantum} {Chemistry}},
	url = {https://pubs.acs.org/doi/10.1021/acs.jctc.2c00802},
	doi = {10.1021/acs.jctc.2c00802},
	language = {en},
	number = {12},
	urldate = {2023-09-04},
	journal = {Journal of Chemical Theory and Computation},
	author = {Lee, Joonho and Pham, Hung Q. and Reichman, David R.},
	month = dec,
	year = {2022},
	pages = {7024--7042},
}

@article{shee2021revealing,
  title={Revealing the nature of electron correlation in transition metal complexes with symmetry breaking and chemical intuition},
  author={Shee, James and Loipersberger, Matthias and Hait, Diptarka and Lee, Joonho and Head-Gordon, Martin},
  journal={The Journal of chemical physics},
  volume={154},
  number={19},
  year={2021},
  publisher={AIP Publishing}
}

@article{zhang2003quantum,
  title={Quantum {M}onte {C}arlo method using phase-free random walks with {S}later determinants},
  author={Zhang, Shiwei and Krakauer, Henry},
  journal={{P}hysical {R}eview {L}etters },
  volume={90},
  number={13},
  pages={136401},
  year={2003},
  publisher={APS}
}

@article{upadhyay2020role,
  title={The role of high-order electron correlation effects in a model system for non-valence correlation-bound anions},
  author={Upadhyay, Shiv and Dumi, Amanda and Shee, James and Jordan, Kenneth D},
  journal={The {J}ournal of {C}hemical {P}hysics},
  volume={153},
  number={22},
  pages={224118},
  year={2020},
  publisher={AIP Publishing LLC}
}

@article{bartlett2007coupled,
  title={Coupled-cluster theory in quantum chemistry},
  author={Bartlett, Rodney J and Musia{\l}, Monika},
  journal={Reviews of Modern Physics},
  volume={79},
  number={1},
  pages={291},
  year={2007},
  publisher={APS}
}

@article{hoffman2014mechanism,
  title={Mechanism of nitrogen fixation by nitrogenase: the next stage},
  author={Hoffman, Brian M and Lukoyanov, Dmitriy and Yang, Zhi-Yong and Dean, Dennis R and Seefeldt, Lance C},
  journal={Chemical reviews},
  volume={114},
  number={8},
  pages={4041--4062},
  year={2014},
  publisher={ACS Publications}
}

@article{askerka2017o2,
  title={The O2-evolving complex of photosystem II: Recent insights from quantum mechanics/molecular mechanics (QM/MM), extended X-ray absorption fine structure (EXAFS), and femtosecond X-ray crystallography data},
  author={Askerka, Mikhail and Brudvig, Gary W and Batista, Victor S},
  journal={Accounts of chemical research},
  volume={50},
  number={1},
  pages={41--48},
  year={2017},
  publisher={ACS Publications}
}

@article{shi2014symmetry,
  title={Symmetry-projected wave functions in quantum Monte Carlo calculations},
  author={Shi, Hao and Jim{\'e}nez-Hoyos, Carlos A and Rodr{\'\i}guez-Guzm{\'a}n, R and Scuseria, Gustavo E and Zhang, Shiwei},
  journal={Physical Review B},
  volume={89},
  number={12},
  pages={125129},
  year={2014},
  publisher={APS}
}

@article{mayer1984bond,
  title={Bond order and valence: Relations to Mulliken's population analysis},
  author={Mayer, I},
  journal={International journal of quantum chemistry},
  volume={26},
  number={1},
  pages={151--154},
  year={1984},
  publisher={Wiley Online Library}
}

@article{sun2018pyscf,
  title={PySCF: the Python-based simulations of chemistry framework},
  author={Sun, Qiming and Berkelbach, Timothy C and Blunt, Nick S and Booth, George H and Guo, Sheng and Li, Zhendong and Liu, Junzi and McClain, James D and Sayfutyarova, Elvira R and Sharma, Sandeep and others},
  journal={Wiley Interdisciplinary Reviews: Computational Molecular Science},
  volume={8},
  number={1},
  pages={e1340},
  year={2018},
  publisher={Wiley Online Library}
}

@article{qin2016benchmark,
  title={Benchmark study of the two-dimensional Hubbard model with auxiliary-field quantum Monte Carlo method},
  author={Qin, Mingpu and Shi, Hao and Zhang, Shiwei},
  journal={Physical Review B},
  volume={94},
  number={8},
  pages={085103},
  year={2016},
  publisher={APS}
}

@article{danilov2025capturing,
  title={Capturing Strong Correlation in Molecules with Phaseless Auxiliary-Field Quantum Monte Carlo Using Generalized Hartree--Fock Trial Wave Functions},
  author={Danilov, Don and Ganoe, Brad and Munyi, Mark and Shee, James},
  journal={Journal of Chemical Theory and Computation},
  volume={21},
  number={3},
  pages={1136--1152},
  year={2025},
  publisher={ACS Publications}
}

@article{manni2021modeling,
  title={Modeling magnetic interactions in high-valent trinuclear [Mn 3 (IV) O 4] 4+ complexes through highly compressed multi-configurational wave functions},
  author={Manni, Giovanni Li},
  journal={Physical Chemistry Chemical Physics},
  volume={23},
  number={35},
  pages={19766--19780},
  year={2021},
  publisher={Royal Society of Chemistry}
}

@article{li2020compression,
  title={Compression of spin-adapted multiconfigurational wave functions in exchange-coupled polynuclear spin systems},
  author={Li Manni, Giovanni and Dobrautz, Werner and Alavi, Ali},
  journal={Journal of Chemical Theory and Computation},
  volume={16},
  number={4},
  pages={2202--2215},
  year={2020},
  publisher={ACS Publications}
}

@article{li2021resolution,
  title={Resolution of low-energy states in spin-exchange transition-metal clusters: Case study of singlet states in [Fe (III) 4S4] cubanes},
  author={Li Manni, Giovanni and Dobrautz, Werner and Bogdanov, Nikolay A and Guther, Kai and Alavi, Ali},
  journal={The Journal of Physical Chemistry A},
  volume={125},
  number={22},
  pages={4727--4740},
  year={2021},
  publisher={ACS Publications}
}

@article{song2025genetic,
  title={A Genetic Algorithm Approach for Compact Wave Function Representations in Spin-Adapted Bases},
  author={Song, Maru and Li Manni, Giovanni},
  journal={Journal of Chemical Theory and Computation},
  year={2025},
  publisher={ACS Publications}
}

@article{dobrautz2021spin,
  title={Spin-pure stochastic-CASSCF via GUGA-FCIQMC applied to iron--sulfur clusters},
  author={Dobrautz, Werner and Weser, Oskar and Bogdanov, Nikolay A and Alavi, Ali and Li Manni, Giovanni},
  journal={Journal of Chemical Theory and Computation},
  volume={17},
  number={9},
  pages={5684--5703},
  year={2021},
  publisher={ACS Publications}
}

@article{yano2014mn4ca,
  title={Mn4Ca cluster in photosynthesis: where and how water is oxidized to dioxygen},
  author={Yano, Junko and Yachandra, Vittal},
  journal={Chemical reviews},
  volume={114},
  number={8},
  pages={4175--4205},
  year={2014},
  publisher={ACS Publications}
}

@article{siegbahn2013water,
  title={Water oxidation mechanism in photosystem II, including oxidations, proton release pathways, O―O bond formation and O2 release},
  author={Siegbahn, Per EM},
  journal={Biochimica et Biophysica Acta (BBA)-Bioenergetics},
  volume={1827},
  number={8-9},
  pages={1003--1019},
  year={2013},
  publisher={Elsevier}
}

@article{blomberg1997comparative,
  title={A comparative study of high-spin manganese and iron complexes},
  author={Blomberg, Margareta RA and Siegbahn, Per EM},
  journal={Theoretical Chemistry Accounts},
  volume={97},
  number={1},
  pages={72--80},
  year={1997},
  publisher={Springer}
}

@article{anderson1950antiferromagnetism,
  title={Antiferromagnetism. Theory of superexchange interaction},
  author={Anderson, Philip W},
  journal={Physical Review},
  volume={79},
  number={2},
  pages={350},
  year={1950},
  publisher={APS}
}

@article{kanamori1959superexchange,
  title={Superexchange interaction and symmetry properties of electron orbitals},
  author={Kanamori, Junjiro},
  journal={Journal of Physics and Chemistry of Solids},
  volume={10},
  number={2-3},
  pages={87--98},
  year={1959},
  publisher={Elsevier}
}

@article{gibson1966iron,
  title={The iron complex in spinach ferredoxin.},
  author={Gibson, JF and Hall, DO and Thornley, JH and Whatley, FR},
  journal={Proceedings of the National Academy of Sciences},
  volume={56},
  number={3},
  pages={987--990},
  year={1966}
}

@article{brintzinger1966ligand,
  title={On the ligand field of iron in ferredoxin from spinach chloroplasts and related nonheme iron enzymes.},
  author={Brintzinger, Hans and Palmer, Graham and Sands, Richard H},
  journal={Proceedings of the National Academy of Sciences},
  volume={55},
  number={2},
  pages={397--404},
  year={1966}
}

@article{dunham1971structure,
  title={On the structure of the iron-sulfur complex in the two-iron ferredoxins},
  author={Dunham, William Richard and Palmer, Graham and Sands, Richard H and Bearden, Alan J},
  journal={Biochimica et Biophysica Acta (BBA)-Bioenergetics},
  volume={253},
  number={2},
  pages={373--384},
  year={1971},
  publisher={Elsevier}
}

@article{bertini1991proton,
  title={Proton NMR spectroscopy and the electronic structure of the high potential iron-sulfur protein from Chromatium vinosum},
  author={Bertini, Ivano and Briganti, Fabrizio and Luchinat, Claudio and Scozzafava, Andrea and Sola, Marco},
  journal={Journal of the American Chemical Society},
  volume={113},
  number={4},
  pages={1237--1245},
  year={1991},
  publisher={ACS Publications}
}

@article{noodleman1986ligand,
  title={Ligand spin polarization and antiferromagnetic coupling in transition metal dimers},
  author={Noodleman, Louis and Davidson, Ernest R},
  journal={Chemical physics},
  volume={109},
  number={1},
  pages={131--143},
  year={1986},
  publisher={Elsevier}
}

@article{blondin1990interplay,
  title={Interplay of electron exchange and electron transfer in metal polynuclear complexes in proteins or chemical models},
  author={Blondin, Genevieve and Girerd, Jean Jacques},
  journal={Chemical Reviews},
  volume={90},
  number={8},
  pages={1359--1376},
  year={1990},
  publisher={ACS Publications}
}

@article{papaefthymiou1987moessbauer,
  title={Moessbauer study of D. gigas ferredoxin II and spin-coupling model for Fe3S4 cluster with valence delocalization},
  author={Papaefthymiou, VGJJ and Girerd, JJ and Moura, I and Moura, JJG and Muenck, E.+},
  journal={Journal of the American Chemical Society},
  volume={109},
  number={15},
  pages={4703--4710},
  year={1987},
  publisher={ACS Publications}
}

@article{noodleman1995orbital,
  title={Orbital interactions, electron delocalization and spin coupling in iron-sulfur clusters},
  author={Noodleman, L and Peng, CY and Case, DA and Mouesca, J-M},
  journal={Coordination Chemistry Reviews},
  volume={144},
  pages={199--244},
  year={1995},
  publisher={Elsevier}
}

@article{noodleman1984electronic,
  title={Electronic structure, magnetic properties, ESR, and optical spectra for 2-iron ferredoxin models by LCAO-X. alpha. valence bond theory},
  author={Noodleman, Louis and Baerends, Evert Jan},
  journal={Journal of the American Chemical Society},
  volume={106},
  number={8},
  pages={2316--2327},
  year={1984},
  publisher={ACS Publications}
}

@article{goodenough1968spin,
  title={Spin-orbit-coupling effects in transition-metal compounds},
  author={Goodenough, John B},
  journal={Physical Review},
  volume={171},
  number={2},
  pages={466},
  year={1968},
  publisher={APS}
}

@article{anderson1959new,
  title={New approach to the theory of superexchange interactions},
  author={Anderson, Philip W},
  journal={Physical Review},
  volume={115},
  number={1},
  pages={2},
  year={1959},
  publisher={APS}
}

@article{kubin2018probing,
  title={Probing the oxidation state of transition metal complexes: a case study on how charge and spin densities determine Mn L-edge X-ray absorption energies},
  author={Kubin, Markus and Guo, Meiyuan and Kroll, Thomas and L{\"o}chel, Heike and K{\"a}llman, Erik and Baker, Michael L and Mitzner, Rolf and Gul, Sheraz and Kern, Jan and F{\"o}hlisch, Alexander and others},
  journal={Chemical science},
  volume={9},
  number={33},
  pages={6813--6829},
  year={2018},
  publisher={Royal Society of Chemistry}
}

@article{mayer1983towards,
  title={Towards a “chemical” Hamiltonian},
  author={Mayer, I},
  journal={International Journal of Quantum Chemistry},
  volume={23},
  number={2},
  pages={341--363},
  year={1983},
  publisher={Wiley Online Library}
}

@article{blomberg2014quantum,
  title={Quantum chemical studies of mechanisms for metalloenzymes},
  author={Blomberg, Margareta RA and Borowski, Tomasz and Himo, Fahmi and Liao, Rong-Zhen and Siegbahn, Per EM},
  journal={Chemical reviews},
  volume={114},
  number={7},
  pages={3601--3658},
  year={2014},
  publisher={ACS Publications}
}

@article{siegbahn2019mechanism,
  title={The mechanism for nitrogenase including all steps},
  author={Siegbahn, Per EM},
  journal={Physical Chemistry Chemical Physics},
  volume={21},
  number={28},
  pages={15747--15759},
  year={2019},
  publisher={Royal Society of Chemistry}
}

@article{cao2019extremely,
  title={Extremely large differences in DFT energies for nitrogenase models},
  author={Cao, Lili and Ryde, Ulf},
  journal={Physical Chemistry Chemical Physics},
  volume={21},
  number={5},
  pages={2480--2488},
  year={2019},
  publisher={Royal Society of Chemistry}
}

@article{sproviero2008computational,
  title={Computational studies of the O2-evolving complex of photosystem II and biomimetic oxomanganese complexes},
  author={Sproviero, Eduardo M and Gasc{\'o}n, Jos{\'e} A and McEvoy, James P and Brudvig, Gary W and Batista, Victor S},
  journal={Coordination chemistry reviews},
  volume={252},
  number={3-4},
  pages={395--415},
  year={2008},
  publisher={Elsevier}
}

@article{cox2013biological,
  title={Biological water oxidation},
  author={Cox, Nicholas and Pantazis, Dimitrios A and Neese, Frank and Lubitz, Wolfgang},
  journal={Accounts of chemical research},
  volume={46},
  number={7},
  pages={1588--1596},
  year={2013},
  publisher={ACS Publications}
}

@article{drosou2024combined,
  title={Combined Multireference--Multiscale Approach to the Description of Photosynthetic Reaction Centers},
  author={Drosou, Maria and Bhattacharjee, Sinjini and Pantazis, Dimitrios A},
  journal={Journal of Chemical Theory and Computation},
  volume={20},
  number={16},
  pages={7210--7226},
  year={2024},
  publisher={ACS Publications}
}

@article{krewald2013magnetic,
  title={On the magnetic and spectroscopic properties of high-valent Mn3CaO4 cubanes as structural units of natural and artificial water-oxidizing catalysts},
  author={Krewald, Vera and Neese, Frank and Pantazis, Dimitrios A},
  journal={Journal of the American Chemical Society},
  volume={135},
  number={15},
  pages={5726--5739},
  year={2013},
  publisher={ACS Publications}
}

@article{krewald2015metal,
  title={Metal oxidation states in biological water splitting},
  author={Krewald, Vera and Retegan, Marius and Cox, Nicholas and Messinger, Johannes and Lubitz, Wolfgang and DeBeer, Serena and Neese, Frank and Pantazis, Dimitrios A},
  journal={Chemical Science},
  volume={6},
  number={3},
  pages={1676--1695},
  year={2015},
  publisher={Royal Society of Chemistry}
}

@article{benediktsson2022analysis,
  title={Analysis of the geometric and electronic structure of spin-coupled iron--sulfur dimers with broken-symmetry DFT: Implications for FeMoco},
  author={Benediktsson, Bardi and Bjornsson, Ragnar},
  journal={Journal of Chemical Theory and Computation},
  volume={18},
  number={3},
  pages={1437--1457},
  year={2022},
  publisher={ACS Publications}
}

@incollection{noodleman1992density,
  title={Density-functional theory of spin polarization and spin coupling in iron—sulfur clusters},
  author={Noodleman, Louis and Case, David A},
  booktitle={Advances in Inorganic Chemistry},
  volume={38},
  pages={423--470},
  year={1992},
  publisher={Elsevier}
}

@article{zhai2023multireference,
  title={Multireference protonation energetics of a dimeric model of nitrogenase iron--sulfur clusters},
  author={Zhai, Huanchen and Lee, Seunghoon and Cui, Zhi-Hao and Cao, Lili and Ryde, Ulf and Chan, Garnet Kin-Lic},
  journal={The Journal of Physical Chemistry A},
  volume={127},
  number={47},
  pages={9974--9984},
  year={2023},
  publisher={ACS Publications}
}

@article{kurashige2013entangled,
  title={Entangled quantum electronic wavefunctions of the Mn4CaO5 cluster in photosystem II},
  author={Kurashige, Yuki and Chan, Garnet Kin-Lic and Yanai, Takeshi},
  journal={Nature chemistry},
  volume={5},
  number={8},
  pages={660--666},
  year={2013},
  publisher={Nature Publishing Group UK London}
}

@article{leyser2024restricted,
  title={Restricted Open-Shell Hartree--Fock Method for a General Configuration State Function Featuring Arbitrarily Complex Spin-Couplings},
  author={Leyser da Costa Gouveia, Tiago and Maganas, Dimitrios and Neese, Frank},
  journal={The Journal of Physical Chemistry A},
  volume={128},
  number={25},
  pages={5041--5053},
  year={2024},
  publisher={ACS Publications}
}

@article{zhai2026classical,
  title={Classical solution of the FeMo-cofactor model to chemical accuracy and its implications},
  author={Zhai, Huanchen and Li, Chenghan and Zhang, Xing and Li, Zhendong and Lee, Seunghoon and Chan, Garnet Kin},
  journal={arXiv preprint arXiv:2601.04621},
  year={2026}
}

@article{shi2021some,
  title={Some recent developments in auxiliary-field quantum Monte Carlo for real materials},
  author={Shi, Hao and Zhang, Shiwei},
  journal={The Journal of chemical physics},
  volume={154},
  number={2},
  year={2021},
  publisher={AIP Publishing}
}

@article{awasthi2025noncovalent,
  title={Noncovalent interaction energies with phaseless auxiliary-field quantum Monte Carlo},
  author={Awasthi, Devesh and Otis, Leon and Huang, Emily and Shee, James},
  journal={Journal of Chemical Theory and Computation},
  year={2025},
  publisher={ACS Publications}
}

@article{vuong2026evaluating,
  title={Evaluating Multiconfigurational Trials for Accurate Phaseless Auxiliary-Field Quantum Monte Carlo on 3d Transition Metal Complexes},
  author={Vuong, Hung T and Mahajan, Ankit and Weber, John L and Shee, James and Reichman, David R and Friesner, Richard A},
  journal={Journal of Chemical Theory and Computation},
  year={2026},
  publisher={ACS Publications}
}

@article{huang2024gpu,
  title={GPU-accelerated Auxiliary-field quantum Monte Carlo with multi-Slater determinant trial states},
  author={Huang, Yifei and Guo, Zhen and Pham, Hung Q and Lv, Dingshun},
  journal={arXiv preprint arXiv:2406.08314},
  year={2024}
}

@article{kjonstad2025systematic,
  title={Systematic improvement of trial states in phaseless auxiliary-field quantum Monte Carlo},
  author={Kj{\o}nstad, Eirik F and Damour, Yann and Sharma, Sandeep and Chan, Garnet Kin},
  journal={arXiv preprint arXiv:2510.06486},
  year={2025}
}

@article{hait2018delocalization,
  title={Delocalization errors in density functional theory are essentially quadratic in fractional occupation number},
  author={Hait, Diptarka and Head-Gordon, Martin},
  journal={The journal of physical chemistry letters},
  volume={9},
  number={21},
  pages={6280--6288},
  year={2018},
  publisher={ACS Publications}
}

@article{Holm1974,
author = {Holm, R. H. and Averill, B. A. and Herskovitz, T. and Frankel, R. B. and Gray, H. B. and Siiman, O. and Grunthaner, F. J.},
title = {Equivalence of metal centers in the iron-sulfur protein active site analogs [Fe4S4(SR)4]2-},
journal = {Journal of the American Chemical Society},
volume = {96},
number = {8},
pages = {2644-2646},
year = {1974},
doi = {10.1021/ja00815a071},
URL = { https://doi.org/10.1021/ja00815a071},
eprint = {https://doi.org/10.1021/ja00815a071}

}

@article{Girerd1989,
url = {https://doi.org/10.1351/pac198961050805},
title = {Double exchange in iron-sulfur clusters and a proposed spin-dependent transfer mechanism},
title = {},
author = {J.-J. Girerd and V. Papaefthymiou and K. K. Surerus and E. Munck},
pages = {805--816},
volume = {61},
number = {5},
journal = {Pure and Applied Chemistry},
doi = {doi:10.1351/pac198961050805},
year = {1989},
lastchecked = {2026-02-23}
}

@article{Noodleman1991a,
author = {Noodleman, Louis},
title = {Exchange coupling and resonance delocalization in reduced iron-sulfur [Fe4S4]+ and iron-selenium [Fe4Se4]+ clusters. 1. Basic theory of spin-state energies and EPR and hyperfine properties},
journal = {Inorganic Chemistry},
volume = {30},
number = {2},
pages = {246-256},
year = {1991},
doi = {10.1021/ic00002a019},
URL = {https://doi.org/10.1021/ic00002a019},
eprint = {https://doi.org/10.1021/ic00002a019}

}

@article{Noodleman1991b,
author = {Noodleman, Louis},
title = {Exchange coupling and resonance delocalization in reduced iron-sulfur [Fe4S4]+ and iron-selenium [Fe4Se4]+ clusters. 2. A generalized nonlinear model for spin-state energies and EPR and hyperfine properties},
journal = {Inorganic Chemistry},
volume = {30},
number = {2},
pages = {256-264},
year = {1991},
doi = {10.1021/ic00002a020},
URL = {https://doi.org/10.1021/ic00002a020},
eprint = {https://doi.org/10.1021/ic00002a020}

}

@article{Papaefthymiou1982,
author = {Papaefthymiou, G. C. and Laskowski, E. J. and Frota-Pessoa, S. and Frankel, R. B. and Holm, R. H.},
title = {Antiferromagnetic exchange interactions in [Fe4S4(SR)4]2-,3- clusters},
journal = {Inorganic Chemistry},
volume = {21},
number = {5},
pages = {1723-1728},
year = {1982},
doi = {10.1021/ic00135a005},
URL = {https://doi.org/10.1021/ic00135a005},
eprint = {https://doi.org/10.1021/ic00135a005}
}

@article{Zener1951,
  title = {Interaction Between the $d$ Shells in the Transition Metals},
  author = {Zener, C.},
  journal = {Phys. Rev.},
  volume = {81},
  issue = {3},
  pages = {440--444},
  numpages = {0},
  year = {1951},
  month = {Feb},
  publisher = {American Physical Society},
  doi = {10.1103/PhysRev.81.440},
  url = {https://link.aps.org/doi/10.1103/PhysRev.81.440}
}

@article{Anderson1955,
  title = {Considerations on Double Exchange},
  author = {Anderson, P. W. and Hasegawa, H.},
  journal = {Phys. Rev.},
  volume = {100},
  issue = {2},
  pages = {675--681},
  numpages = {0},
  year = {1955},
  month = {Oct},
  publisher = {American Physical Society},
  doi = {10.1103/PhysRev.100.675},
  url = {https://link.aps.org/doi/10.1103/PhysRev.100.675}
}

@article{Holm1977,
author = {Holm, Richard H.},
title = {Synthetic approaches to the active sites of iron-sulfur proteins},
journal = {Accounts of Chemical Research},
volume = {10},
number = {12},
pages = {427-434},
year = {1977},
doi = {10.1021/ar50120a001},
URL = {https://doi.org/10.1021/ar50120a001
},
eprint = {https://doi.org/10.1021/ar50120a001}

}

@article{Rao2004,
author = {Venkateswara Rao, P. and Holm, R. H.},
title = {Synthetic Analogues of the Active Sites of Iron-Sulfur Proteins},
journal = {Chemical Reviews},
volume = {104},
number = {2},
pages = {527-560},
year = {2004},
doi = {10.1021/cr020615+},
URL = {https://doi.org/10.1021/cr020615+},
eprint = {https://doi.org/10.1021/cr020615+}
}

@article{Palmer1966,
title = {Electron paramagnetic resonance studies on the ferredoxin from Clostridium pasteurianum},
journal = {Biochemical and Biophysical Research Communications},
volume = {23},
number = {4},
pages = {357-362},
year = {1966},
issn = {0006-291X},
doi = {https://doi.org/10.1016/0006-291X(66)90733-9},
url = {https://www.sciencedirect.com/science/article/pii/0006291X66907339},
author = {Graham Palmer and Richard H. Sands and L.E. Mortenson}
}

@article{Blomstrom1964,
author = {D. C. Blomstrom  and E. Knight  and W. D. Phillips  and J. F. Weiher },
title = {The Nature of Iron in Ferredoxin},
journal = {Proceedings of the National Academy of Sciences},
volume = {51},
number = {6},
pages = {1085-1092},
year = {1964},
doi = {10.1073/pnas.51.6.1085},
URL = {https://www.pnas.org/doi/abs/10.1073/pnas.51.6.1085},
eprint = {https://www.pnas.org/doi/pdf/10.1073/pnas.51.6.1085}}

@article{Poe1970,
author = {M. Poe  and W. D. Phillips  and C. C. McDonald  and W. Lovenberg },
title = {Proton Magnetic Resonance Study of Ferredoxin from Clostridium pasteurianum},
journal = {Proceedings of the National Academy of Sciences},
volume = {65},
number = {4},
pages = {797-804},
year = {1970},
doi = {10.1073/pnas.65.4.797},
URL = {https://www.pnas.org/doi/abs/10.1073/pnas.65.4.797},
eprint = {https://www.pnas.org/doi/pdf/10.1073/pnas.65.4.797}}

@article{jiang2024improved,
  title={Improved modularity and new features in ipie: Toward even larger AFQMC calculations on CPUs and GPUs at zero and finite temperatures},
  author={Jiang, Tong and Baumgarten, Moritz KA and Loos, Pierre-Fran{\c{c}}ois and Mahajan, Ankit and Scemama, Anthony and Ung, Shu Fay and Zhang, Jinghong and Malone, Fionn D and Lee, Joonho},
  journal={The Journal of Chemical Physics},
  volume={161},
  number={16},
  year={2024},
  publisher={AIP Publishing}
}

@article{al2007bond,
  title={Bond breaking with auxiliary-field quantum Monte Carlo},
  author={Al-Saidi, Wissam A and Zhang, Shiwei and Krakauer, Henry},
  journal={The Journal of chemical physics},
  volume={127},
  number={14},
  year={2007},
  publisher={AIP Publishing}
}

@article{motta2017towards,
  title={Towards the solution of the many-electron problem in real materials: Equation of state of the hydrogen chain with state-of-the-art many-body methods},
  author={Motta, Mario and Ceperley, David M and Chan, Garnet Kin-Lic and Gomez, John A and Gull, Emanuel and Guo, Sheng and Jim{\'e}nez-Hoyos, Carlos A and Lan, Tran Nguyen and Li, Jia and Ma, Fengjie and others},
  journal={Physical Review X},
  volume={7},
  number={3},
  pages={031059},
  year={2017},
  publisher={APS}
}

@article{zhang1995constrained,
  title={Constrained path quantum Monte Carlo method for fermion ground states},
  author={Zhang, Shiwei and Carlson, J and Gubernatis, James E},
  journal={Physical review letters},
  volume={74},
  number={18},
  pages={3652},
  year={1995},
  publisher={APS}
}

@article{huang2024evaluating,
  title={Evaluating a quantum-classical quantum Monte Carlo algorithm with Matchgate shadows},
  author={Huang, Benchen and Chen, Yi-Ting and Gupt, Brajesh and Suchara, Martin and Tran, Anh and McArdle, Sam and Galli, Giulia},
  journal={Physical Review Research},
  volume={6},
  number={4},
  pages={043063},
  year={2024},
  publisher={APS}
}

@article{zhao2025quantum,
  title={Quantum-classical auxiliary field quantum Monte Carlo with matchgate shadows on trapped ion quantum computers},
  author={Zhao, Luning and Goings, Joshua J and Aboumrad, Willie and Arrasmith, Andrew and Calderin, Lazaro and Churchill, Spencer and Gabay, Dor and Harvey-Brown, Thea and Hiles, Melanie and Kaja, Magda and others},
  journal={arXiv preprint arXiv:2506.22408},
  year={2025}
}

@article{yoshida2025auxiliary,
  title={Auxiliary-field quantum Monte Carlo method with quantum selected configuration interaction},
  author={Yoshida, Yuichiro and Erhart, Luca and Murokoshi, Takuma and Nakagawa, Rika and Mori, Chihiro and Miyanaga, Takafumi and Mori, Toshio and Mizukami, Wataru},
  journal={arXiv preprint arXiv:2502.21081},
  year={2025}
}

@article{amsler2023classical,
  title={Classical and quantum trial wave functions in auxiliary-field quantum Monte Carlo applied to oxygen allotropes and a CuBr2 model system},
  author={Amsler, Maximilian and Deglmann, Peter and Degroote, Matthias and Kaicher, Michael P and Kiser, Matthew and K{\"u}hn, Michael and Kumar, Chandan and Maier, Andreas and Samsonidze, Georgy and Schroeder, Anna and others},
  journal={The Journal of Chemical Physics},
  volume={159},
  number={4},
  year={2023},
  publisher={AIP Publishing}
}

@article{mejuto2022effect,
  title={The effect of geometry, spin, and orbital optimization in achieving accurate, correlated results for iron--sulfur cubanes},
  author={Mejuto-Zaera, Carlos and Tzeli, Demeter and Williams-Young, David and Tubman, Norm M and Matousek, Mikulas and Brabec, Jiri and Veis, Libor and Xantheas, Sotiris S and de Jong, Wibe A},
  journal={Journal of chemical theory and computation},
  volume={18},
  number={2},
  pages={687--702},
  year={2022},
  publisher={ACS Publications}
}

@article{skeel2025iron,
  title={Iron-sulfur clusters: the road to room temperature},
  author={Skeel, Brighton A and Suess, Daniel LM},
  journal={JBIC Journal of Biological Inorganic Chemistry},
  volume={30},
  number={2},
  pages={151--159},
  year={2025},
  publisher={Springer}
}

@article{qin2016coupling,
  title={Coupling quantum Monte Carlo and independent-particle calculations: Self-consistent constraint for the sign problem based on the density or the density matrix},
  author={Qin, Mingpu and Shi, Hao and Zhang, Shiwei},
  journal={Physical Review B},
  volume={94},
  number={23},
  pages={235119},
  year={2016},
  publisher={APS}
}

@article{sukurma2025self,
  title={Self-refinement of auxiliary-field quantum Monte Carlo via non-orthogonal configuration interaction},
  author={Sukurma, Zoran and Schlipf, Martin and Kresse, Georg},
  journal={Journal of Chemical Theory and Computation},
  volume={21},
  number={9},
  pages={4481--4493},
  year={2025},
  publisher={ACS Publications}
}

@article{ung2026study,
  title={Study of the triangular-lattice Hubbard model with constrained-path quantum Monte Carlo},
  author={Ung, Shu Fay and Mahajan, Ankit and Reichman, David R},
  journal={arXiv preprint arXiv:2603.14808},
  year={2026}
}

\end{document}


\maketitle

\begin{figure}[H]
    \centering
    \includegraphics[width=0.75\linewidth]{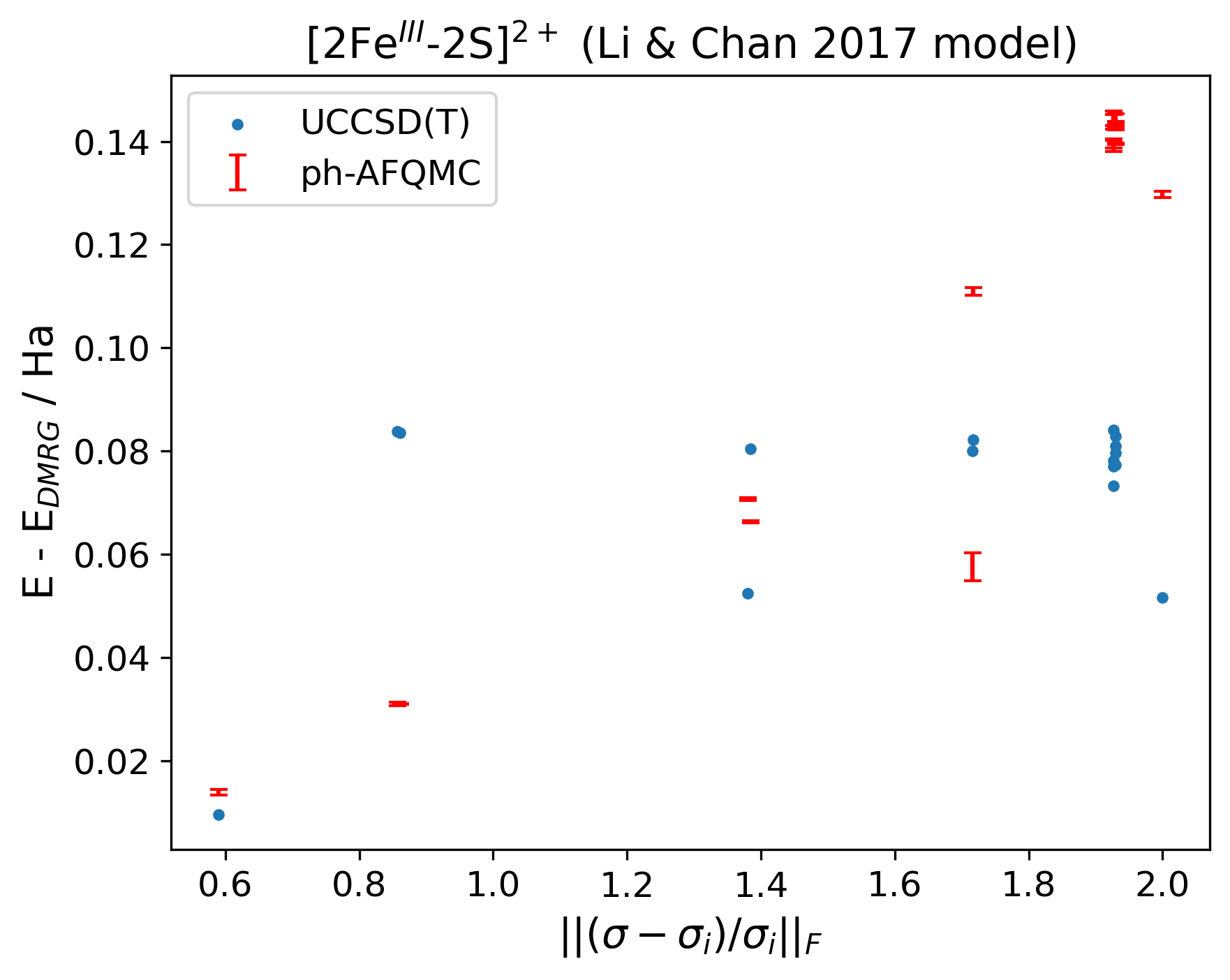}
    \caption{\footnotesize Deviations of the ph-AFQMC-UHF and UCCSD(T) energies from the DMRG reference, as a function of the SDM of the UHF trials.}
    \label{fig:2FeSDM}
\end{figure}

\

\section{SCF, UCCSD(T), and ph-AFQMC data for the three iron-sulfur clusters}

We note that a subset of the UHF states that we found either led to numerical issues or pathological energy vs imaginary-time trajectories that prevented us from obtaining meaningful ph-AFQMC energy estimates.  For [4Fe--4S]$^{2+}$ and [4Fe--4S]$^{4+}$ there were 9 and 2 such cases, respectively.  The CCSD(T) energies corresponding to these references are nevertheless shown in the Figures in the main text.

\clearpage
\begin{landscape}
\subsection{[2Fe$^{III}$ - 2S]$^{2+}$}

\begin{table}[]
\begin{tabular}{rr
>{\columncolor[HTML]{F4CCCC}}r 
>{\columncolor[HTML]{F4CCCC}}r 
>{\columncolor[HTML]{F4CCCC}}r 
>{\columncolor[HTML]{FCE5CD}}l 
>{\columncolor[HTML]{FCE5CD}}l 
>{\columncolor[HTML]{CFE2F3}}l 
>{\columncolor[HTML]{CFE2F3}}l }
\multicolumn{1}{l}{} & \multicolumn{1}{l}{} & \multicolumn{3}{l}{\cellcolor[HTML]{F4CCCC}Spin density analysis} & \multicolumn{2}{l}{\cellcolor[HTML]{FCE5CD}Unrestricted coupled cluster} & \multicolumn{2}{l}{\cellcolor[HTML]{CFE2F3}Phaseless AFQMC} \\
\multicolumn{1}{l}{E UHF} & \multicolumn{1}{l}{S\textasciicircum{}2 UHF} & \multicolumn{1}{l}{\cellcolor[HTML]{F4CCCC}SD Fe I} & \multicolumn{1}{l}{\cellcolor[HTML]{F4CCCC}SD Fe II} & \multicolumn{1}{l}{\cellcolor[HTML]{F4CCCC}SDM} & E UCCSD & E UCCSD(T) & E UHF-ph-AFQMC & Err UHF-ph-AFQMC \\
\cellcolor[HTML]{FFFFFF}-116.512 692 & \cellcolor[HTML]{FFFFFF}4.8932 & 4.1977 & -4.2004 & 0.5894 & -116.580 470 & -116.596 038 & -116.591 769 & 0.000 555 \\
\cellcolor[HTML]{FFFFFF}-116.488 952 & \cellcolor[HTML]{FFFFFF}4.8234 & 3.7785 & -3.7821 & 0.8567 & -116.518 797 & -116.521 819 & -116.574 570 & 0.000 342 \\
\cellcolor[HTML]{FFFFFF}-116.487 806 & \cellcolor[HTML]{FFFFFF}4.8213 & 3.7745 & -3.7724 & 0.8609 & -116.519 615 & -116.522 021 & -116.574 624 & 0.000 086 \\
\cellcolor[HTML]{FFFFFF}-116.428 922 & \cellcolor[HTML]{FFFFFF}4.5850 & 3.7395 & -1.9684 & 1.3798 & -116.534 919 & -116.553 129 & -116.534 927 & 0.000 322 \\
\cellcolor[HTML]{FFFFFF}-116.428 164 & \cellcolor[HTML]{FFFFFF}4.5763 & 3.7375 & -1.9519 & 1.3843 & -116.520 370 & -116.525 138 & -116.539 292 & 0.000 226 \\
\cellcolor[HTML]{FFFFFF}-116.367 533 & \cellcolor[HTML]{FFFFFF}4.3867 & -1.8834 & 1.8863 & 1.7158 & -116.518 475 & -116.525 637 & -116.548 033 & 0.002 711 \\
\cellcolor[HTML]{FFFFFF}-116.367 071 & \cellcolor[HTML]{FFFFFF}4.3853 & 1.8751 & -1.8851 & 1.7172 & -116.516 220 & -116.523 398 & -116.494 710 & 0.000 706 \\
\cellcolor[HTML]{FFFFFF}-116.347 679 & \cellcolor[HTML]{FFFFFF}4.4623 & -0.0778 & 1.8883 & 1.9264 & -116.508 619 & -116.521 588 & -116.466 533 & 0.001 078 \\
\cellcolor[HTML]{FFFFFF}-116.347 452 & \cellcolor[HTML]{FFFFFF}4.4610 & 1.8874 & -0.0784 & 1.9265 & -116.515 457 & -116.528 626 & -116.465 979 & 0.000 905 \\
\cellcolor[HTML]{FFFFFF}-116.347 448 & \cellcolor[HTML]{FFFFFF}4.4769 & 0.0790 & -1.8878 & 1.9264 & -116.514 030 & -116.527 475 & -116.461 759 & 0.001 428 \\
\cellcolor[HTML]{FFFFFF}-116.347 216 & \cellcolor[HTML]{FFFFFF}4.4757 & -1.8860 & 0.0785 & 1.9265 & -116.517 888 & -116.532 319 & -116.461 135 & 0.001 446 \\
\cellcolor[HTML]{FFFFFF}-116.346 817 & \cellcolor[HTML]{FFFFFF}4.4123 & 0.0400 & -1.8915 & 1.9293 & -116.516 572 & -116.528 367 & -116.461 242 & 0.000 980 \\
\cellcolor[HTML]{FFFFFF}-116.346 573 & \cellcolor[HTML]{FFFFFF}4.4125 & -1.8872 & 0.0369 & 1.9298 & -116.512 302 & -116.524 657 & -116.466 090 & 0.000 145 \\
\cellcolor[HTML]{FFFFFF}-116.346 093 & \cellcolor[HTML]{FFFFFF}4.4192 & 1.8914 & -0.0343 & 1.9297 & -116.513 649 & -116.525 975 & -116.462 049 & 0.000 336 \\
\cellcolor[HTML]{FFFFFF}-116.346 091 & \cellcolor[HTML]{FFFFFF}4.4175 & 0.0331 & -1.8901 & 1.9299 & -116.511 286 & -116.522 770 & -116.463 139 & 0.000 201 \\
\cellcolor[HTML]{FFFFFF}-116.338 560 & \cellcolor[HTML]{FFFFFF}4.4426 & -0.0068 & 0.0045 & 2.0000 & -116.539 236 & -116.554 002 & -116.475 891 & 0.000 576
\end{tabular}
\end{table}
\end{landscape}
\newpage

\subsection{[4Fe-4S]$^{2+}$}

\begin{table}[]
\small
\begin{tabular}{ll
>{\columncolor[HTML]{F4CCCC}}l 
>{\columncolor[HTML]{F4CCCC}}l 
>{\columncolor[HTML]{F4CCCC}}l 
>{\columncolor[HTML]{F4CCCC}}l 
>{\columncolor[HTML]{F4CCCC}}l 
>{\columncolor[HTML]{F4CCCC}}l 
>{\columncolor[HTML]{F4CCCC}}l }
             &                          & \multicolumn{7}{c}{\cellcolor[HTML]{F4CCCC}UHF spin density analysis}                                                                 \\
E UHF        & S$^2$ UHF & SD Fe 1 & SD Fe 2 & SD Fe 3 & SD Fe 4 & {[++-- --]}    & {[+--+--]}          & {[+-- --+]}          \\
-327.083 894 & 8.8259                   & 3.7435  & -3.8017 & 4.0549  & -3.8692 & 4.9737                        & 1.0534                        & 4.9729                        \\
-327.083 452 & 8.7977                   & 3.9684  & -3.7418 & 3.7452  & -3.9831 & 4.9652                        & 1.0649                        & 4.9673                        \\
-327.079 922 & 8.3167                   & 3.6839  & -3.6838 & 3.6801  & -3.6823 & 4.8141                        & 1.3213                        & 4.8141                        \\
-327.079 094 & 8.8742                   & 3.8752  & -3.8827 & 3.8897  & -3.8770 & 4.9853                        & 1.0245                        & 4.9853                        \\
-327.021 907 & 8.3717                   & 3.7269  & -2.0797 & 3.7239  & -3.9838 & 4.6096                        & 1.8893                        & 4.6099                        \\
-326.958 138 & 8.2026                   & 3.7283  & -4.0091 & 1.9947  & -1.8676 & 4.4048                        & 2.4757                        & 4.2350                        \\
-326.945 905 & 8.1112                   & 0.0916  & -3.7186 & 3.9570  & -1.8628 & 3.9265                        & 3.0807                        & 4.2721                        \\
-326.945 887 & 8.1054                   & 3.9497  & -1.8628 & 0.0945  & -3.7190 & 3.9268                        & 3.0805                        & 4.2716                        \\
-326.942 477 & 7.9985                   & 3.6842  & -3.7907 & -0.0742 & 1.8184  & 4.2478                        & 3.9121                        & 3.1170                        \\
-326.940 586 & 8.0501                   & 3.8016  & -3.6957 & 0.0745  & -1.8884 & 4.2525                        & 3.0998                        & 3.9272                        \\
-326.940 343 & 7.9921                   & 1.8693  & 3.6951  & -0.0786 & -3.7941 & 3.1034                        & 4.2514                        & 3.9236                        \\
-326.928 622 & 7.9896                   & 3.6743  & -3.6749 & -0.0130 & 0.0132  & 4.2165                        & 3.5330                        & 3.5222                        \\
-326.927 039 & 7.9665                   & -0.021  & 3.6764  & 0.0207  & -3.6767 & 3.5359                        & 4.2169                        & 3.5187                        \\
-326.923 198 & 7.8585                   & -3.7015 & -0.0745 & 0.0864  & 3.6910  & 3.4911                        & 3.5577                        & 4.2217                        \\
-326.923 009 & 7.8577                   & -0.0821 & -3.6931 & 3.6979  & 0.0767  & 3.4916                        & 3.5574                        & 4.2215                        \\
-327.021 412 & 8.4063                   & 3.962   & -3.6649 & 1.9543  & -3.7039 & 4.5713                        & 1.9697                        & 4.5747                        \\
-327.021 919 & 8.4112                   & 1.9497  & -3.7034 & 3.9731  & -3.6663 & 4.5726                        & 1.9692                        & 4.5759                        \\
-327.079 407 & 8.3244                   & 3.6764  & 3.6908  & -3.6752 & -3.6912 & 1.3200                        & 4.8148                        & 4.8148                        \\
-327.082 038 & 8.8309                   & -4.0545 & -3.7454 & 3.8688  & 3.8128  & 1.0484                        & 4.9764                        & 4.9757                        \\
-327.082 766 & 8.7828                   & -4.0528 & -3.7271 & 3.7504  & 3.9163  & 1.0640                        & 4.9672                        & 4.9694                        \\
-327.082 836 & 8.8121                   & -1.9229 & 3.7086  & -4.0382 & 3.6814  & 4.5831                        & 1.9630                        & 4.5856                        \\
-327.083 187 & 8.8254                   & -4.0526 & 3.8715  & -3.7433 & 3.8005  & 4.9734                        & 1.0538                        & 4.9726                        \\
-327.083 499 & 8.7959                   & -3.7548 & -3.9218 & 4.0542  & 3.7323  & 1.0576                        & 4.9709                        & 4.9731                        \\
-327.083 505 & 8.8068                   & -3.7511 & -3.9687 & 3.9847  & 3.7482  & 1.0592                        & 4.9684                        & 4.9705                        \\
-327.084 217 & 8.8293                   & -3.8044 & -3.8773 & 3.7454  & 4.0554  & 1.0482                        & 4.9768                        & 4.9759                        \\
-327.084 303 & 8.8020                   & -3.7451 & 3.9736  & -3.9817 & 3.7461  & 4.9670                        & 1.0618                        & 4.9692                        \\
-327.084 355 & 8.8292                   & -3.7456 & -4.0555 & 3.8042  & 3.8768  & 1.0484                        & 4.9767                        & 4.9758                        \\
-327.089 868 & 8.7449                   & -3.7229 & 3.9534  & 3.7219  & -3.9518 & {\color[HTML]{333333} 4.9453} & {\color[HTML]{333333} 4.9474} & {\color[HTML]{333333} 1.0980} \\
-327.080 211 & 8.4512                   & 3.6592  & -3.6598 & -3.7389 & 3.7388  & {\color[HTML]{333333} 4.8278} & {\color[HTML]{333333} 4.8276} & {\color[HTML]{333333} 1.2977} \\
-327.082 733 & 8.8226                   & 3.7966  & -3.7410 & 3.8770  & -4.0534 & {\color[HTML]{333333} 4.9735} & {\color[HTML]{333333} 1.0539} & {\color[HTML]{333333} 4.9725} \\
-327.082 759 & 8.8323                   & 3.8640  & 3.8158  & -4.0575 & -3.7461 & {\color[HTML]{333333} 1.0478} & {\color[HTML]{333333} 4.9768} & {\color[HTML]{333333} 4.9762} \\
-327.082 796 & 8.8234                   & -3.8736 & 4.0541  & -3.8006 & 3.7418  & {\color[HTML]{333333} 4.9740} & {\color[HTML]{333333} 1.0531} & {\color[HTML]{333333} 4.9730} \\
-327.082 953 & 8.8050                   & 3.9685  & 3.7509  & -3.7477 & -3.9802 & {\color[HTML]{333333} 1.0611} & {\color[HTML]{333333} 4.9672} & {\color[HTML]{333333} 4.9692} \\
-327.088 412 & 8.7434                   & -3.9496 & 3.9473  & 3.7219  & -3.7221 & {\color[HTML]{333333} 4.9454} & {\color[HTML]{333333} 4.9433} & {\color[HTML]{333333} 1.1012} \\
-327.088 874 & 8.7440                   & -3.7233 & 3.7209  & 3.9548  & -3.9431 & {\color[HTML]{333333} 4.9456} & {\color[HTML]{333333} 4.9436} & {\color[HTML]{333333} 1.1008} \\
-326.949 870 & 7.9336                   & -3.6803 & 1.8109  & -1.8162 & 3.6909  & {\color[HTML]{333333} 4.1704} & {\color[HTML]{333333} 2.6134} & {\color[HTML]{333333} 4.3332} \\
-326.949 382 & 8.2010                   & 4.0371  & -3.6977 & -1.7969 & 0.0350  & 4.2723                        & 3.9047                        & 3.1063                        \\
-326.938 715 & 8.2131                   & 3.7263  & 1.7593  & -0.0105 & -4.0115 & 3.1208                        & 3.8922                        & 4.2730                        \\
-326.923 857 & 7.8544                   & 3.6879  & -0.0726 & 0.0882  & -3.7018 & 3.5579                        & 3.4913                        & 4.2213                        \\
-327.017 894 & 8.4304                   & 3.7123  & 3.6783  & -1.9243 & -4.0385 & 1.9622                        & 4.5832                        & 4.5863                        \\
-327.018 985 & 8.4339                   & 3.7106  & -1.9320 & 3.6841  & -4.0372 & 4.5852                        & 1.9580                        & 4.5876                        \\
-327.019 836 & 8.4342                   & -1.9229 & 3.7086  & -4.0382 & 3.6814  & 4.5831                        & 1.9630                        & 4.5856                        \\
-327.080 834 & 8.4634                   & -3.6659 & 3.7357  & 3.6636  & -3.7360 & 4.8285                        & 4.8287                        & 1.2959                        \\
-327.088 966 & 8.7404                   & -3.9408 & 3.7212  & 3.9476  & -3.7227 & 4.9414                        & 4.9434                        & 1.1043                        \\
-327.019 003 & 8.4324                   & 4.0362  & -3.6799 & 1.9220  & -3.7092 & 4.5824                        & 1.9640                        & 4.5851                       
\end{tabular}
\end{table}

\begin{table}[]
\small
\begin{tabular}{l
>{\columncolor[HTML]{FCE5CD}}l 
>{\columncolor[HTML]{FCE5CD}}l 
>{\columncolor[HTML]{CFE2F3}}l 
>{\columncolor[HTML]{CFE2F3}}l }
             & \multicolumn{2}{c}{\cellcolor[HTML]{FCE5CD}Unrestricted coupled cluster} & \multicolumn{2}{c}{\cellcolor[HTML]{CFE2F3}Phaseless AFQMC} \\
E UHF        & E UCCSD                             & E UCCSD(T)                         & E UHF-ph-AFQMC              & Err UHF-ph-AFQMC              \\
-327.083 894 & -327.196 581                        & -327.215 198                       & -327.227 380                & 0.000 387                     \\
-327.083 452 & -327.199 640                        & -327.215 887                       & -327.232 177                & 0.000 280                     \\
-327.079 922 & -327.173 533                        & -327.184 365                       & -327.222 583                & 0.000 579                     \\
-327.079 094 & -327.209 282                        & -327.221 853                       & -327.245 651                & 0.000 266                     \\
-327.021 907 & -327.160 757                        & -327.176 295                       & -327.163 374                & 0.000 540                     \\
-326.958 138 & -327.172 088                        & -327.223 862                       & -327.122 142                & 0.000 313                     \\
-326.945 905 & -327.162 136                        & -327.190 365                       & -327.117 978                & 0.000 297                     \\
-326.945 887 & -327.159 212                        & -327.186 143                       & -327.118 176                & 0.000 428                     \\
-326.942 477 & -327.144 400                        & -327.168 719                       & -327.110 665                & 0.000 419                     \\
-326.940 586 & -327.154 684                        & -327.185 571                       & -327.118 406                & 0.000 221                     \\
-326.940 343 & -327.151 845                        & -327.180 514                       & -327.115 770                & 0.000 264                     \\
-326.928 622 & -327.152 484                        & -327.181 056                       & -327.091 647                & 0.001 791                     \\
-326.927 039 & -327.167 122                        & -327.215 791                       & -327.086 861                & 0.001 504                     \\
-326.923 198 & -327.139 332                        & -327.164 584                       & -327.094 702                & 0.000 292                     \\
-326.923 009 & -327.134 836                        & -327.159 445                       & -327.093 844                & 0.000 388                     \\
-327.021 412 & -327.173 917                        & -327.195 878                       & -327.183 357                & 0.000 520                     \\
-327.021 919 & -327.174 316                        & -327.196 056                       & -327.182 087                & 0.000 238                     \\
-327.079 407 & -327.173 790                        & -327.184 311                       & -327.223 261                & 0.000 196                     \\
-327.082 038 & -327.195 569                        & -327.214 465                       & -327.226 065                & 0.000 330                     \\
-327.082 766 & -327.197 087                        & -327.215 601                       & -327.223 718                & 0.000 403                     \\
-327.082 836 & -327.169 595                        & -327.193 200                       & -327.229 044                & 0.000 272                     \\
-327.083 187 & -327.196 116                        & -327.214 925                       & -327.227 074                & 0.000 358                     \\
-327.083 499 & -327.197 464                        & -327.216 086                       & -327.225 334                & 0.000 614                     \\
-327.083 505 & -327.199 391                        & -327.215 846                       & -327.233 029                & 0.000 442                     \\
-327.084 217 & -327.197 123                        & -327.215 908                       & -327.227 401                & 0.000 311                     \\
-327.084 303 & -327.200 126                        & -327.216 412                       & -327.233 989                & 0.000 354                     \\
-327.084 355 & -327.197 167                        & -327.215 913                       & -327.227 691                & 0.000 239                     \\
-327.089 868 & -327.199 165                        & -327.213 268                       & -327.222 101                & 0.001 294                     \\
-327.080 211 & -327.192 248                        & -327.209 510                       & -327.217 698                & 0.000 175                     \\
-327.082 733 & -327.195 510                        & -327.214 554                       & -327.234 881                & 0.000 587                     \\
-327.082 759 & -327.196 285                        & -327.214 864                       & -327.233 998                & 0.000 679                     \\
-327.082 796 & -327.196 028                        & -327.215 000                       & -327.231 198                & 0.000 438                     \\
-327.082 953 & -327.199 448                        & -327.215 859                       & -327.231 848                & 0.000 295                     \\
-327.088 412 & -327.197 841                        & -327.211 844                       & -327.221 230                & 0.000 554                     \\
-327.088 874 & -327.198 034                        & -327.212 002                       & -327.221 637                & 0.000 795                     \\
-326.949 870 & -327.141 710                        & -327.164 312                       & *                           & *                             \\
-326.949 382 & -327.112 415                        & -327.140 216                       & *                           & *                             \\
-326.938 715 & -327.178 672                        & -327.218 410                       & *                           & *                             \\
-326.923 857 & -327.131 461                        & -327.153 431                       & *                           & *                             \\
-327.017 894 & -327.168 280                        & -327.191 340                       & *                           & *                             \\
-327.018 985 & -327.166 236                        & -327.188 156                       & *                           & *                             \\
-327.019 836 & -327.169 595                        & -327.193 200                       & *                           & *                             \\
-327.080 834 & -327.191 994                        & -327.209 548                       & *                           & *                             \\
-327.088 966 & -327.198 048                        & -327.211 723                       & *                           & *                             \\
-327.019 003 & *                                   & *                                  & *                           & *                            
\end{tabular}
\end{table}

\restoregeometry
\clearpage

\begin{landscape}
\subsection{[4Fe$^{III}$ - 4S]$^{4+}$}
\begin{table}[]
\begin{tabular}{lr
>{\columncolor[HTML]{F4CCCC}}r 
>{\columncolor[HTML]{F4CCCC}}r 
>{\columncolor[HTML]{F4CCCC}}r 
>{\columncolor[HTML]{F4CCCC}}r 
>{\columncolor[HTML]{F4CCCC}}r 
>{\columncolor[HTML]{F4CCCC}}r 
>{\columncolor[HTML]{F4CCCC}}r }
 & \multicolumn{1}{l}{} & \multicolumn{7}{c}{\cellcolor[HTML]{F4CCCC}UHF spin density analysis} \\
\multicolumn{1}{c}{E UHF} & \multicolumn{1}{c}{S\textasciicircum{}2 UHF} & \multicolumn{1}{c}{\cellcolor[HTML]{F4CCCC}SD Fe 1} & \multicolumn{1}{c}{\cellcolor[HTML]{F4CCCC}SD Fe 2} & \multicolumn{1}{c}{\cellcolor[HTML]{F4CCCC}SD Fe 3} & \multicolumn{1}{c}{\cellcolor[HTML]{F4CCCC}SD Fe 4} & \multicolumn{1}{c}{\cellcolor[HTML]{F4CCCC}SDM [+ + -- --]} & \multicolumn{1}{c}{\cellcolor[HTML]{F4CCCC}SDM [+ -- + --]} & \multicolumn{1}{c}{\cellcolor[HTML]{F4CCCC}SDM [+ -- -- +]} \\
-8432.351 925 & 8.7921 & 1.7408 & -3.6351 & -3.6352 & 4.3074 & 4.3774 & 4.3774 & 2.3447 \\
-8432.351 982 & 8.8424 & -4.2991 & 3.6387 & -1.7725 & 3.6418 & 4.3811 & 2.3332 & 4.3814 \\
-8432.352 255 & 8.8465 & 1.7758 & -3.6372 & 4.2997 & -3.6458 & 4.3824 & 2.3314 & 4.3816 \\
-8432.436 186 & 9.2833 & 4.3014 & 3.6349 & -4.3024 & -3.6398 & 1.5087 & 4.7377 & 4.7228 \\
-8432.438 961 & 9.2845 & -3.9873 & 3.6364 & -3.6379 & 3.9709 & 4.6226 & 1.6862 & 4.6266 \\
-8432.439 381 & 9.2841 & 4.2984 & 3.6372 & -3.6372 & -4.3008 & 1.5100 & 4.7220 & 4.7368 \\
-8432.440 251 & 9.2545 & -3.3309 & 3.6352 & 4.2919 & -3.3131 & 4.5364 & 4.5368 & 1.9052 \\
-8432.487 411 & 9.7777 & -4.2926 & 4.2928 & 4.2931 & -4.2946 & 4.9697 & 4.9697 & 1.0508 \\
-8432.488 177 & 9.8325 & -4.2992 & 4.2994 & -4.2999 & 4.3015 & 4.9752 & 1.0416 & 4.9752 \\
-8432.488 458 & 9.8329 & -4.2991 & -4.2993 & 4.3000 & 4.3016 & 1.0416 & 4.9752 & 4.9752
\end{tabular}
\end{table}
\end{landscape}

\begin{landscape}
    
\begin{table}[]
\begin{tabular}{l
>{\columncolor[HTML]{FCE5CD}}l 
>{\columncolor[HTML]{FCE5CD}}l 
>{\columncolor[HTML]{CFE2F3}}l 
>{\columncolor[HTML]{CFE2F3}}l }
                          & \multicolumn{2}{c}{\cellcolor[HTML]{FCE5CD}Unrestricted coupled cluster}                                     & \multicolumn{2}{c}{\cellcolor[HTML]{CFE2F3}Phaseless AFQMC}                                                               \\
\multicolumn{1}{c}{E UHF} & \multicolumn{1}{c}{\cellcolor[HTML]{FCE5CD}E UCCSD} & \multicolumn{1}{c}{\cellcolor[HTML]{FCE5CD}E UCCSD(T)} & \multicolumn{1}{c}{\cellcolor[HTML]{CFE2F3}E UHF-ph-AFQMC} & \multicolumn{1}{c}{\cellcolor[HTML]{CFE2F3}Err UHF-ph-AFQMC} \\
-8432.351 925             & -8432.535 560                                       & -8432.563 949                                          & -8432.549 679                                              & 0.000 220                                                    \\
-8432.351 982             & -8432.532 618                                       & -8432.563 949                                          & -8432.541 112                                              & 0.000 279                                                    \\
-8432.352 255             & -8432.533 364                                       & -8432.567 478                                          & -8432.542 492                                              & 0.000 147                                                    \\
-8432.436 186             & -8432.562 568                                       & -8432.581 664                                          & -8432.610 029                                              & 0.000 300                                                    \\
-8432.438 961             & -8432.565 246                                       & -8432.583 300                                          &                                                            &                                                              \\
-8432.439 381             & -8432.565 738                                       & -8432.583 826                                          & -8432.606 266                                              & 0.000 152                                                    \\
-8432.440 251             & -8432.572 221                                       & -8432.591 322                                          & -8432.610 427                                              & 0.000 534                                                    \\
-8432.487 411             & -8432.602 227                                       & -8432.625 151                                          & -8432.637 557                                              & 0.003 400                                                    \\
-8432.488 177             & -8432.594 458                                       & -8432.615 520                                          & -8432.642 349                                              & 0.000 418                                                    \\
-8432.488 458             & -8432.488 458                                       & -8432.615 597                                          &                                                            &                                                             
\end{tabular}
\end{table}

\end{landscape}

\clearpage
